\documentclass[]{aa}
\usepackage[dvips]{graphicx}
\usepackage{amsmath}

\newcommand{\mincir}{\raise
  -2.truept\hbox{\rlap{\hbox{$\sim$}}\raise5.truept \hbox{$<$}\ }}
\newcommand{\magcir}{\raise
  -2.truept\hbox{\rlap{\hbox{$\sim$}}\raise5.truept \hbox{$>$}\ }}                            

\begin{document}

\title{Heating groups and clusters of galaxies: the role of AGN jets}
\author{Claudio Zanni\inst{1}, Giuseppe Murante\inst{2}, 
	Gianluigi Bodo\inst{2}, 
	Silvano Massaglia\inst{1}, 
	Paola Rossi\inst{2}, 
      Attilio Ferrari\inst{1}
	}
\authorrunning{Zanni et al.}
\institute{Dipartimento di
Fisica Generale, Universit\'a degli Studi di Torino, Torino (Italy) 
\and INAF, Osservatorio Astronomico di Torino, Pino Torinese (Italy)}

\abstract{
X-Ray observations of groups and clusters of galaxies show that the
Intra-Cluster Medium (ICM) in their cores is hotter than expected from 
cosmological numerical simulations of cluster formation which include star
formation, radiative cooling and SN feedback.
We investigate the effect of the injection of supersonic AGN jets into 
the ICM using axisymmetric hydrodynamical numerical simulations. 
A simple model for the ICM, describing the radial properties of
gas and the gravitational potential in cosmological 
N-Body+SPH simulations of one cluster and three groups of galaxies at
redshift $z=0$, is obtained and used to set the environment in which
the jets are injected. We varied the kinetic power of the jet
and the emission-weighted X-Ray temperature of the ICM. The jets
transfer their energy to the ICM mainly by the effects of their
terminal shocks. A high fraction of the injected energy can be
deposited through irreversible processes in the cluster gas, 
up to 75\% in our simulations. We show
how one single, powerful jet can reconcile the predicted X-Ray properties of small groups,  
e.g. the $L_X-T_X$ relation, with observations. We argue that
the interaction between AGN jets and galaxy groups and cluster
atmospheres is a viable feedback mechanism.
}
\maketitle

\section{Introduction}
\label{sec:intro}
 
Clusters and groups of galaxies are observed in the X-ray band mainly due to the
thermal bremsstrahlung radiation emitted by the hot gas, 
the Intra Cluster Medium (ICM). A wealth of recent observations, coming
in particular from the X--ray telescopes CHANDRA and XMM-Newton (Paerels \& Kahan 2003),
have evidenced a complex ICM structure, which is characterized not only by
cooling cores (Peterson et al. 2003, Kaastra et al. 2004), 
but also by cold fronts, bow--shocks, ``bullet--like''
structures (Vikhlinin et al. 2001, Markevitch et al. 2002, Kempner et al. 2003). 
While all of these features can be understood in terms of
the currently favored hierarchical model of structure formation,
there is an ongoing debate concerning the origin and the details of the
energy budget of the ICM. 

In fact, the most obvious way to explain the heating of the ICM to the
temperatures observed in groups and clusters of galaxies, ranging from 
$\approx 0.5$ to more than $10$ keV, is to suppose a pure
gravitational heating. Such a heating is a natural outcome of the
merging processes which form groups and clusters of galaxies in the
hierarchical model picture. While the bigger and bigger Dark Matter
halos merge, their gas is shocked and heats up, to
form the groups and clusters atmospheres which are observed.
The power spectrum of the initial
density perturbations grow hierarchically to form groups and
clusters and does not possess a characteristic length or mass scale, in the
relevant range. The gravity itself has no characteristic
scale. Thus, with the additional assumption that gas is in
hydrostatic equilibrium with the cluster potential, it can be shown
that this scenario predicts simple self--similar scaling relations
among the various observable X--ray properties of groups and
clusters. In particular, the relation between X--ray luminosity and
the temperature is $L_X \propto T^{2}$, and the entropy $S =T/n_e^{2/3}$
(where $n_e$ is the electron number density of the gas) scale as
$S \propto T$ when $S$ is evaluated at fixed overdensities for
different clusters.

Nevertheless, the observational situation appears to be quite different. The
luminosity--temperature relation is steeper than this simple picture
predicts, $L_X \propto T^{\alpha}$ with $\alpha \simeq 2.5-3$ for
clusters with  temperature $T > 2$ keV (White, Jones \& Forman 1997;
Markevitch 1998; Arnaud \& Evrard 1999; Ettori, De Grandi \& Molendi 2002).
There are indications of a steeper slope for groups
with $T<1$keV (Ponman et al. 1996, Helsdon \& Ponman 2000, Sanderson
et al. 2003).
The gas entropy in clusters is found to be in excess with respect to
the model predictions (Ponman, Cannon \& Navarro 1999; Lloyd-Davies, Ponman \& Cannon 2000;
Finoguenov et al. 2003), and its dependence on
the temperature approximately follows the law $S \propto T^{2/3}$
(Ponman, Sanderson \& Finoguenov 2003).

The discrepancy between observations and theory suggests that the ICM is
subject to a heating which is larger than what the simple
self--similar model predicts. 
In general, there is a consensus on the fact that the observations
can be matched using models where a given amount of energy per
particle is added to that coming from gravitational processes,
independently on the precise origin of this energy. 
Numerical studies suggest that a  heating energy of about $0.5-1$ keV 
per particle, or equivalently an entropy floor
of $\approx 50-100$ keV cm$^2$, is enough to satisfy the observational
constraints (Bialek et al. 2001, Brighenti \& Mathews 2001, Borgani et
al. 2001, 2002). Semi--analytical models give an even higher estimate
of the needed heating, up to $\approx 400$ keV cm$^2$ (Babul et al. 2002,
McCarthy, Babul \& Balogh 2002).  
The ICM can be heated by two natural processes, which are present when
the baryons physics is considered. The first is the energy
feedback from supernovae explosions, which directly heats cold gas and
increases its entropy. The second is the radiative cooling
of the gas itself, which subtracts low--entropy cold gas from the
hot phase and turn it into stars, thus increasing the average 
temperature and entropy of the remaining gaseous medium. 
The efficiency of the two processes in heating the ICM have been
extensively studied so far, both with semi--analytical techniques
(Cavaliere, Menci \& Tozzi 1998; Tozzi \& Norman 2001; Menci \&
Cavaliere 2000; Bower et al. 2001; Voit \& Bryan 2001; Voit et
al. 2002; Wu \& Xue 2002) and
with numerical simulations (Evrard \& Henry 1991; Kaiser 1991; Bower
1997; Balog et al. 1999; Muanwong et al. 2002; Dav\`e et al. 2002;
Babul et al. 2002). 

Recent numerical studies also considered simultaneously the effect of
radiative cooling, star formation and feedback from SN (Suginohara \&
Ostriker 1998; Lewis et al. 2000; Yoshida et al. 2002; Loken et
al. 2002; Tornatore et al. 2003; Borgani et al. 2004). These studies
partially disagree on
the resulting agreement with observations which can be obtained,
ranging from a more (Loken et al. 2002) to a less (Borgani et al. 2004) 
optimistic interpretation of the simulations results. All of them,
however, showed how the fraction $f_*$ of gas which is converted
into a ``cold'' stellar phase results to be too high, in excess with
respect to the value $f_* \mincir 10\%$ which is obtained
from measurements of the local luminosity density of stars (Balog et
al. 2001; Lin, Mohr \& Stanford 2003). The overcooling problem which
naturally descend from these results still calls for the need of
further  energy contributions to the ICM to be solved.

A possible source for this energy is given by AGNs activity. The
available energy budget is in principle more than adequate to heat the
ICM to the desired temperature (Valageas \& Silk 1999, Cavaliere, Lapi \& Menci
2002). In fact, the accretion of gas onto supermassive BHs in galactic
cores gives outputs of order $2 \cdot 10^{62} (M/10^9 M_\odot)$ erg, where
$M$ is the accreted mass and a standard mass--energy conversion
efficiency of $10^{-1}$ is used (Wu, Fabian \& Nulsen 2000; Bower et
al. 2001). 
Observations with the ROSAT satellite suggested an interaction
of the central AGN activity with the ICM in the Perseus cluster 
(B\"oringer et al. 1993), possibly under the form of ``bubbles''
(McNamara, O'Connel \& Sarazin 1996). 
CHANDRA also gave observational hints of 
possible coupling between ICM and central radio--sources in clusters:
observations of the Perseus cluster by Fabian et al. (2000) showed that the cluster
gas is displaced by the radio lobes of the source 3C84 which is inflating cavities
or ``bubbles'' in the X-ray emitting gas; observations of ripples and sound waves 
propagating through the ICM (Fabian et al. 2003a) suggest that the cavities are gently 
expanding and interacting with the surrounding medium. A similar scenario hes been depicted 
in the case of M87 by Forman et al. (2003) who observed weak shock fronts associated with
the AGN outburst propagating through the surrounding gas. 
In the case of the Hydra A cluster, hosting the powerful
FRI radio source 3C218, it has been argued that, even if there are no signs of shock--heated gas
around the radio lobes (McNamara et al. 2000), shock heating is needed in the cluster core 
to balance cooling flows (David et al. 2001). 
Recent deep CHANDRA observations of the same source (Nulsen et al. 2004) 
have revealed features in the X-Ray bightness profile which can be
associated with a shock front driven by the expanding radio lobes requiring
a total energy around $10^{61}$ erg to drive it.
Observations of Cygnus A (Smith et al. 2002) have revealed signs of stronger interaction
between the radio source and the ICM since the shells of enhanced X-ray emission around
the cavity are hotter than the surrounding medium perhaps as a result of heating by a bow--shock
driven by the expanding cavity. 
X-ray emission from shocked gas has been detected in the observations of the radio galaxy 
PKS 1138-262 at z=2.156 (Carilli et al. 2002) for which a jet kinetic power $5 \times 10^{46}$ erg s$^{-1}$ 
injected for $2\times 10^7$ years has been estimated. 
The X-ray emission observed around the FRII radio source 3C294 at z=1.786 (Fabian et al. 2003),
if interpreted as thermal radiation from shocked material, gives an estimates of the injected power
greater than $10^{46}$ erg s$^{-1}$.  

On the other hand, we still miss a coherent and physically
well--based model of the coupling of this energy with the
ICM. Different studies of coupling between the AGN power and the ICM
have been done, which can be roughly divided in two main classes: the
study of the buoyant behavior of hot gas bubbles in the ICM,
almost independently of how the bubbles were produced (Quilis,
Bower \& Balogh 2001; Churazov et al. 2001, 2002; Br\"uggen et
al. 2002; Br\"uggen \& Kaiser 2002; De Young 2002; McCarthy et
al. 2003; Mathews et al. 2003;
Hoeft,  Br\"uggen \& Yepes 2003), and the direct study of the effect of
injection of mechanical energy, via an AGN jet, in the ICM (Cavaliere,
Lapi \& Menci 2002; Reynolds, Heinz \& Begelman 2002; Nath \&
Roychowdhury 2002). In the first view, the energy supply to the
cluster is gentle and buoyancy--driven; in the second view, it is a
momentum--driven process, characterized by the formation and
dissipation of shock fronts. Other ideas have been proposed as well,
for instance the viscous dissipation of sound waves (Pringle 1989)
which can arise from the activity of a central AGN (Ruszkowski,
Br\"uggen \&  Begelman 2003).

Semi--analytical studies of the above processes are usually based on simple
prescription for the exchange of energy between the central AGN and the
ICM, coupled with basic models for describing the dynamical evolution
of bubbles, cocoons or jets  (Cavaliere, Lapi \& Menci 2002; Nath \&
Roychowdhury 2002; Churazov et al. 2002; De Young 2002; McCarthy et
al. 2003; Mathews et al. 2003) and, at best, a modeling of the
cluster atmosphere (Wu, Fabian \& Nulsen 2000), 
usually consisting either in a classical
$\beta$--model (Cavaliere \& Fusco-Femiano 1976) or in a modified
model of the same kind, in which the hot cluster gas in in hydrostatic
equilibrium with the gravitational potential of the galaxy cluster. In
the last case, the mass density profile of the cluster is usually taken to be a
Navarro, Frenk \& White (NFW) analytical profile, which have been shown to
correctly describe, {\it on average}, the shape of such clusters in
cosmological N--body simulations (Navarro, Frenk \& White
1996, 1997). The gravitational potential is then derived consequently.

Hydrodynamical numerical simulations of the ICM heating by AGN also
resort  to these class of models for the cluster atmosphere,
sometimes slightly modified for closely following an observed
distribution of the mass and electron density temperature (Nulsen \&
B\"{o}hringher 1995). Simulations of buoyant bubbles are usually
performed in three dimensions, at the cost of a lower resolution
(Quilis, Bower \& Balogh 2001; Br\"uggen et al. 2002) 
or in two dimension assuming spherical symmetry thus reaching good 
resolution only in the central part of the computational domain 
(Churazov et al. 2001). 

Simulations of jets in the ICM have been performed in 2D with spherical 
symmetry by Reynolds, Heinz \& Begelman (2002).
Their jet was injected in a cluster atmosphere for $\approx 5
\cdot 10^7$ yr. The atmosphere was taken to be a simple $\beta$--model,
and a suitable static gravitational potential was added to guarantee
the hydrostatic equilibrium. No attempt has been done there to model
the differences which can arise in the ICM as a function
of the group/cluster mass, nor to have a mass profile similar to the
one coming from cosmological simulations.
The atmosphere was then evolved for as long as
$10^9$ yr.  The main result of this work is that, when an equilibrium
is achieved again by the cluster atmosphere, its core specific entropy
has been enhanced by $\approx 20$ per cent and a large fraction of the
injected energy, as high as $\approx 50$ per cent, is thermalized onto
the ICM. A three dimensional study of the interaction between a low
power jet and the ICM has been performed by Omma et al. (2004) having
a resolution around $\approx$ 0.6 kpc with 1024 cells per side.

A numerical study of viscous dissipation of waves had still higher
length resolution ($\approx 0.1$ kpc with 2048 cells per side, Ruszkowski,
Br\"uggen \&  Begelman 2003). They showed how the viscous
dissipation of the energy of sound--waves produced by a single AGN
duty--cycle is insufficient to balance radiative cooling of the gas;
but intermittent activity of the central source is able to offset
cooling by this process. 

Finally, a cosmological self--consistent study of the evolution of the
radio plasma during a merger of cluster of galaxies has been performed
by Hoeft, Br\"uggen \& Yepes (2003). They used the Tree+SPH N-body
code GADGET (Springel, Yoshida \& White 2001) and achieved a high mass and
force resolution (the mass of a gas particles was $M_{gas}=2.6 \cdot
10^7 h^{-1} M_\odot$ and the Plummer--equivalent softening length was
$\epsilon = 2 h^{-1}$ kpc). However, their focus was on the evolution
of the fossil radio--relics and not on the energy exchanges between
the AGN which produced such relics and the ICM.

In the present work, we study the energy exchange between AGN--powered jets
and the surrounding ICM of group/cluster of galaxies, with 2D 
hydrodynamical numerical simulations. We extracted the groups and
clusters atmospheres from  cosmological numerical
simulations (Tornatore et al. 2003), at the redshift $z=0$. We build a
self--similar model for such atmospheres based on the NFW radial
density profile  for the mass distribution and the hydrostatic
equilibrium  with the
consequent potential for the gas. We note that our model is not an
{\it average} description of a typical group or cluster of galaxy, 
but it has been built to represent the output of a self--consistent
numerical cosmological simulation. 
The jet is injected in the ICM for a time $t
\approx 2\cdot 10^7$ yr, then switched off. We varied the temperature
scale of our model and the kinetic energy of the jet. We derive the
energy balance of the ICM, the entropy evolution, and, for the first
time using this approach,  the resulting $L_X-T_X$ relation, after the
heating of the ICM has occurred. 

The plan of the paper is the following: in Section \ref{sec:model}, we
describe the baseline model. In Section \ref{sec:sim} the numerical
simulations are presented. Section \ref{sec:ene} is devoted to the
calculation of the energy balance and Section \ref{sec:entropy} to the entropy
evolution. In Section \ref{sec:equilibrium} we calculate a new hydrostatic equilibrium
for the ICM and derive the $L_X-T_X$, $L_X-M_{200}$ and $S-T_X$ relations. 
In Section \ref{sec:concl} we draw our conclusions and discuss future perspectives of this work.

\section{The baseline model}
\label{sec:model}

A self-consistent, numerical study of the interplay between AGN jets
and clusters atmospheres would require a huge dynamical range, to
resolve simultaneously the kpc scale of the jets and the $\approx$ 100
Mpc scale
required for following the clusters formation and evolution. Such a
dynamical range is still outside the current computing power, even for
supercomputers. Moreover, the details of the formation and evolution
of the AGN itself are not completely understood, thus they cannot
be parametrized easily in self-consistent, cosmological numerical simulations.
To achieve high spatial resolution, we perform 2D hydrodynamical
simulations in cylindrical symmetry. Therefore, we cannot directly use the N-Body+SPH
simulated cluster atmospheres as initial conditions: being 3D, their
resolution is significantly lower than the one we are going to reach. 
We thus build a simple model for the cluster atmospheres, to reproduce
the spherically averaged properties of the cluster gas in cosmological Tree+SPH simulations.
We consider adiabatic
simulations performed by Tornatore et al. (2003, T03 hereinafter) of the evolution
of four clusters having virial mass $M_{\rm vir} = 2.35,\; 2.52,\; 5.98,\; 39.4 \times
10^{13} \; M_\odot$. We take the clusters  at $z=0$ in order 
to directly estimate the effect of jets energy injection on
the X-ray properties of galaxy clusters and to compare these properties with
observations. The parameters of our model are constrained using the radial profiles
of these simulated objects. An additional requirement is that the model
does scale with cluster mass. Our initial conditions for the
environment in which the jets propagate are then set, in our
hydrodynamical simulation, using such model.
As in T03, we use a cosmology with
$\Omega_m=0.3$, $\Omega_\Lambda = 0.7$, Hubble constant $H_0 = 100 \; h \;
{\rm km \; s}^{-1} \; {\rm Mpc}$ with $h=0.7$, and 
$\Omega_{\rm bar}=0.019 \; h^{-2}$ which gives a cosmic baryon fraction
of $f_{\rm bar} \simeq 0.13$. The baryonic gas is taken with a
primordial composition (hydrogen mass-fraction $X=0.76$ and $Y=0.24$ for
helium, mean molecular weight $\mu = 0.6$).
We define here the virial radius $R_{\rm vir}$ as the radius
of a spherical volume inside which the mean density is $\Delta_{\rm c}(z)$
times the critical density ($M_{\rm vir}(z)=4\pi
R_{\rm vir}^3\rho_{\rm crit }\Delta_{\rm c}/3 $),
where $\Delta_{\rm c}(z)$ is the solution of
the spherical collapse of a top-hat perturbation in an expanding
universe (Bryan \& Norman, 1998). At $z=0$, for our chosen cosmology
is $\Delta_{\rm c}=101$.

Firstly we model the profiles of the dark matter density taken from
the T03 simulations with a NFW profile
(Navarro, Frenk \& White 1996):    
\begin{equation}
\label{eq:darkm}
\rho_{\rm dm}=\frac{\rho_{\rm dm0}}{x(1+x)^2}
\end{equation}
with $x=r/r_{\rm s}$ where $r_{\rm s}$ is a characteristic radius.
The dark matter central density $\rho_{\rm dm0}$ is determined by requiring that the dark
matter mass contained inside the virial radius is equal to $(1- f_{\rm bar})M_{\rm vir}$. 
The only free parameter of the dark matter distribution is therefore the concentration 
parameter $c=R_{\rm vir}/r_{\rm s}$ defined by the ratio between the virial radius of the cluster
$R_{\rm vir}$ and the characteristic radius $r_{\rm s}$: from the T03 simulations, we
estimate an average value of $c=9.5$ for our four objects. We use this
value of the concentration parameter for modeling the DM profiles of
every cluster, since we want to obtain a model whose properties do
scale with the virial radius of the objects (via the value of the DM
central density). This assumption is not completely correct, since it
is well known (Navarro, Frenk \& White 1996, 1997; Bullock et
al. 2001) that the concentration of a dark 
matter halo is determined by its mass and its formation history; at a
given mass, early formation
generally leads to a denser core and thus to a higher concentration.
However, taking the average value for $c$ from the Tree+SPH simulations at
$z=0$ we have a maximum error of $10\%$ on the individual values of
the concentration parameter. 

In our calculations, the distribution of the dark matter density determines
the shape of the gravitational potential well. This
potential will be considered steady in time: this approximation 
is reasonable, since the cosmological timescale for the evolution of a
DM halo is much longer than the dynamical timescale of an AGN jet.   
The gravitational potential $\phi$ generated by the density distribution of Eq. 
(\ref{eq:darkm}) is
\begin{equation}
\label{eq:potdm}
\phi=-\frac{G(1-f_{\rm bar})M_{\rm vir}}
{R_{\rm vir}\left[\frac{\log(1+c)}{c}-\frac{1}{1+c}\right]} 
\frac{\left( 1+x \right)}{x}
\end{equation}
The gas distribution is determined imposing hydrostatic equilibrium with
the potential of Eq. (\ref{eq:potdm}) and neglecting the self gravity of the
gas. Assuming a polytropic equation of state for the gas in initial 
equilibrium ($P=P_0(\rho/\rho_0)^\Gamma$) we obtain  
\begin{equation}
\label{eq:rhogas}
\rho = \rho_0 \left\{ \beta \frac{\Gamma-1}{\Gamma}
\left[ \frac{\log(1+x)}{x}-1 \right] +1\right\}^{1/(\Gamma-1)}
\end{equation}
The value taken for the polytropic index is $\Gamma=1.1$, in order to
reproduce the decrease of temperature observed in the T03 simulated
clusters and groups at $z=0$. The coefficient $\beta$ given by the 
expression
\begin{equation}
\label{eq:beta}
\beta = \frac{G(1-f_{\rm bar})M_{\rm vir}}
{R_{\rm vir }\left[\frac{\log(1+c)}{c}-\frac{1}{1+c}\right]}
\frac{\gamma}{c_{\rm s0}^2}
\end{equation} 
is taken constant at all the physical scales in order to
have a self-similar density profile and is equal to $\beta=8$,
which is a good approximation for the T03 simulations.
In Eq. (\ref{eq:beta}) the index $\gamma$ is the ratio of the specific heats
and $c_{\rm s0}=\sqrt{\gamma k T_0/\mu m_{\rm H}}$ is the central
sound speed. Finally the gas central density $\rho_0$ is determined by imposing that the
gas mass inside the virial radius is equal to $f_{\rm bar}M_{\rm vir}$.

With all the above assumptions we obtain the following scaling relations for our
baseline model:
\begin{equation}
\nonumber
n_{\rm e0} = 0.033 \; {\rm cm}^{-3} 
\end{equation}
\[
KT_0 = 7.43 \left( \frac{M_{\rm vir}}{10^{15}M_\odot}\right)^{2/3} \; {\rm keV} 
\]
\[ 
KT_{\rm ew} = 6 \left( \frac{M_{\rm vir}}{10^{15}M_\odot}\right)^{2/3} \; {\rm keV} \; \; \;  
\]
\begin{equation}
\label{eq:scalrel}
KT_{\rm mw} = 4.5 \left( \frac{M_{\rm vir}}{10^{15}M_\odot}\right)^{2/3}\; {\rm keV} 
\end{equation}
\begin{equation}
\nonumber
r_{\rm s} = 272 \left( \frac{M_{\rm vir}}{10^{15}M_\odot}\right)^{1/3} \; {\rm kpc}
\end{equation}
\begin{equation}
\nonumber
L_{\rm br} = 4.34 \times 10^{43}\left( \frac{KT_{\rm ew}}{{\rm keV}} \right)^2 \; {\rm erg \; s}^{-1}   
\end{equation} 
\begin{equation}
\nonumber
S_{0.1 \; R_{200}}=\frac{T}{n_{\rm e}^{2/3}}= 36.3 \left( \frac{KT_{\rm ew}}{\rm keV} \right) \; {\rm keV \; cm}^2
\end{equation}
for, respectively, the central electron density $n_{\rm e0}$, the central temperature
$KT_0$, the emission weighted temperature $KT_{\rm ew}$, the mass weighted 
temperature $KT_{\rm mw}$, the characteristic radius $r_{\rm s}$, the
bremsstrahlung luminosity $L_{\rm br}$ and the entropy $S$ calculated at
$0.1 \; R_{200}$ as in Ponman et al. (2003). 
The emission weighted temperature, assuming bremsstrahlung emissivity, 
and the mass weighted temperature are defined as
\[ 
T_{\rm ew} = 
\frac{\int_0^{R_{\rm vir}}\rho^2T^{3/2}dV}
{\int_0^{R_{\rm vir}}\rho^2T^{1/2}dV} \qquad 
T_{\rm mw}= 
\frac{\int_0^{R_{\rm vir}}\rho T dV}
{\int_0^{R_{\rm vir}}\rho dV} \; . 
\]
Anyway, in the following, we will also include metal--line cooling to compute
the X-Ray luminosities.  
Our assumption of a constant concentration parameter for all clusters
does reflect in a constant central electron density. 
Since our baseline model is self-similar
at all scales, as it is shown by Eqs. (\ref{eq:scalrel}), 
all the characteristic parameters describing the model (central
density, central temperature, virial mass, etc.) scale with only one
free parameter. Such parameter can be chosen as 
the mass $M_{\rm vir}$ or equivalently the virial radius $R_{\rm vir}$, 
the temperature $T_{\rm ew}$, etc., of the cluster or group.

Having derived an analytical expressions for the distribution of all
the physically
interesting quantities and determined the parameters from the T03 
Tree+SPH cosmological simulations,
we can finally test how well the model agrees with the radial
profiles, i.e. how accurately the T03 data are reproduced.

\begin{figure}[ht]
\resizebox{\hsize}{!}{\includegraphics[angle=90]{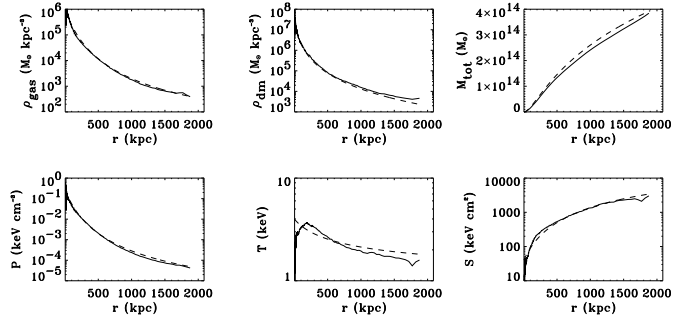}}
\caption{Comparison between our baseline model ({\it dashed line}) and 
and the results of the adiabatic simulation by Tornatore et al. (2003) ({\it solid line}) 
of a cluster with a mass $M_{\rm vir}=39.4 \times 10^{13} \; M_\odot$, taken at $z=0$. 
We show the radial dependencies for the following quantities: 
({\it{Upper panels}}) gas density $\rho$, dark matter density $\rho_{\rm dm}$,
total mass $M_{\rm tot}$; 
({\it{Lower panels}}) gas pressure $P$, gas temperature $T$ and gas entropy
$S=T/n_{\rm e}^{2/3}$ }
\label{fig:cluster}
\end{figure}

\begin{figure}[ht]
\resizebox{\hsize}{!}{\includegraphics[angle=90]{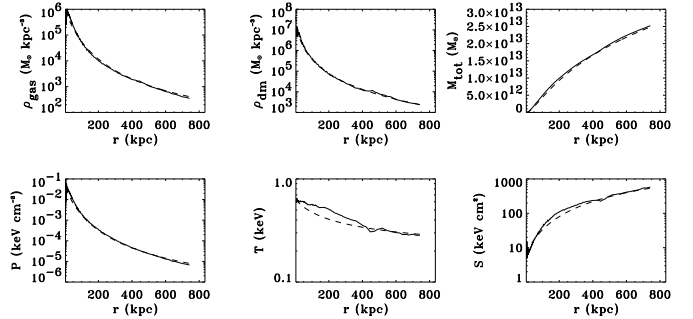}}
\caption{
Same comparison as Fig. \ref{fig:cluster}, 
for a group with a mass 
$M_{\rm vir}=2.52 \times 10^{13} \; M_\odot$. }
\label{fig:group}
\end{figure}

In Figs. \ref{fig:cluster} and \ref{fig:group} we show the comparison between the radial 
dependencies of some quantities as derived from the T03 simulations and as modeled by us for
two clusters of galaxies with respectively a virial mass of $M_{\rm vir} = 39.4 \times 10^{13} \; M_\odot$
and $M_{\rm vir} = 2.52 \times 10^{13} \; M_\odot$.
The agreement of the model with the simulation data appears to be good
for both objects, as far as $\rho$, $\rho_{\rm dm}$, $M_{\rm
Tot}$, $P$, ans $S$ are concerned. There are slight discrepancies only
in the behavior of $T$, especially for the most massive cluster of
Fig. \ref{fig:cluster}, where the central drop is not captured by the
model (see however T03 for a discussion of temperature profiles in SPH simulations).
As a consequence, the analytical model also reproduces the behavior
of the X-ray properties shown by T03 numerical simulations. 
We therefore conclude that the initial
conditions for our hydrodynamical simulations can safely been set
using the model presented in this Section.

\section{The simulations}
\label{sec:sim}
In order to investigate the interaction of an extragalactic jet with the intracluster
gas, we performed two dimensional hydrodynamic numerical simulations of a cylindrical (axisymmetric) jet 
propagating in a gravitationally stratified but not self-gravitating atmosphere:
\begin{eqnarray}
\label{eq:syst}
\nonumber
\frac{\partial \rho}{\partial t}+\nabla\cdot(\rho \vec {v}) & = & 0 \\
\frac{\partial \vec {v}}{\partial t}+(\vec {v} \cdot \nabla)\vec{v}+\nabla
p/\rho & = & \nabla \phi \\
\frac{\partial p}{\partial t}+(\vec{v}\cdot
\nabla)p-\gamma\frac{p}{\rho}\left[\frac{\partial \rho}{\partial
t}+(\vec{v}\cdot\nabla)\rho\right] & =&  0  \; ,
\nonumber
\end{eqnarray}
where the fluid variables $p$, $\rho$ and $\vec{v}$ are
pressure, density and velocity respectively;
$\gamma=5/3$ is the ratio of the specific heats. 
Radiative losses are neglected. The cooling times in the core of the unperturbed 
clusters are longer than the time of activity of our jets (see below). The cooling 
times in the centers are of the same order of magnitude of the elapsed time at the
end of the simulations. On the other hand, the material near the center of the
clusters is subject to significant expansion due to the energy injection itself and
therefore we expect the importance of cooling to be reduced.
The system of Eqs. (\ref{eq:syst}) has been solved numerically employing a PPM 
(Piecewise Parabolic Method) hydrocode (Woodward \& Colella 1984).
As discussed in Sect. \ref{sec:model} the density profile of the 
undisturbed ambient medium is given by Eq. (\ref{eq:rhogas}) while the gravitational
potential $\phi$ needed to take the gas distribution in equilibrium is given by
Eq. (\ref{eq:potdm}).
Simulations have been performed to test the numerical stability of the initial gravitational
equilibrium revealing that small subsonic speeds begin to appear in the central core of the
initial profile on timescales much longer than the energy injection time thus ruling
out the possibility of spurious heating effects.
A (cylindrical) jet is injected from the bottom boundary of the integration domain, 
in pressure balance with the ambient. The axis of the jet is  along the left boundary 
of the domain ($r=0$), where we have imposed  symmetric boundary conditions for
$p, \rho, v_z$ and antisymmetric conditions for $v_r$. Reflective boundary
conditions are also imposed on the boundary of injection of the jet ($z=0$) outside its 
radius in order to reproduce a bipolar flow and to avoid spurious inflow  effects. 
On the remaining outer boundaries we extrapolated the equilibrium
profiles in order to keep the
initial atmosphere in equilibrium on the outer parts of the computational domain.
 
Measuring lengths in units of the characteristic radius $r_{\rm s}$, velocities in units of the
central adiabatic sound speed $c_{\rm s0}$ in the undisturbed external medium and the density in
units of the ambient central density $\rho_0$, our main parameters are the
Mach number $M \equiv v_{\rm j}/c_{\rm s0}$ and the density ratio $\nu = \rho_{\rm j}/
\rho_0$. 
In the following discussion we will take the emission weighted
temperature $T_{\rm ew}$ as
free parameter for describing the unperturbed medium. 
Consistently the unit of time is
\begin{equation}
t_0 = \frac {r_{\rm s}}  {c_{\rm s0}} = 1.9 \times 10^8
\hbox{\rm \; years}
\end{equation}
Defining the units for the jet kinetic power as
\[
L_{\rm k} = \frac {\pi} {2} \rho_0 r_{\rm j}^2 c_{\rm{s}0}^3 
\]
where $r_{\rm j}$ is the jet radius, the kinetic power $L_{\rm j}$
can be expressed as
\begin{equation}
L_{\rm j}=M^3 \nu L_{\rm k} \; .
\label{eq:L_j}
\end{equation}
Assuming a jet radius $r_{\rm j}=4$ kpc $L_{\rm k}$ is given by
\begin{equation}
L_{\rm k} = 3\times 10^{42} \left( \frac {T_{\rm ew}} {{\rm keV}} \right)^{3/2}
{\rm erg \; s^{-1}}
\end{equation}
The cylindrical symmetry adopted in our simulations, contrary to spherical symmetry, allows us to have a high resolution 
from the jet scale up to the cluster scale. We inject a jet having $0^o$ opening angle.  
The assumed jet radius, dictated mostly by the need of an adequate resolution (at least 10 grid points) on the 
jet radius scale, is consistent with the size of jets with an opening angle around $9^o-15^o$ 
(Feretti et al. 1999, Lara et al. 2004) at a distance $\sim 15 - 25$ kpc from the parent core.
Moreover we must consider that the
lifetime of the jets is at least one order of magnitude shorter than the total time followed in the simulations
and after the jet has ceased its activity, as we will show in detail in the next Section, it
loses its collimation and completely disappears: from that time onward the assumed jet radius is no more
important and the whole dynamics of the system is determined, as we will show later, by the total
energy which has been injected. 
In order to follow separately the evolution of the injected material and the ambient medium we define 
a set of four passive tracers following the evolution equation
\[
\frac{\partial \mathcal{T}_{\rm i}}{\partial t}+(\vec{v} \cdot \nabla)\mathcal{T}_{\rm i} = 0
\]
The tracers are initialized as follows: $\mathcal{T}_1=1$ for $r/r_{\rm s} \in \; [0,1]$, $\mathcal{T}_2=1$ for $r/r_{\rm s} \in \; ]1,3]$,
$\mathcal{T}_3=1$ for $r/r_{\rm s} \in \; ]3,6]$, $\mathcal{T}_4=1$ for $r/r_{\rm s} \in \; ]6,9.5]$ and equal to 0 otherwise.      

\begin{table*} 
\caption{Parameters of the six simulations performed: Mach number $M$, 
emission weighted temperature of the ambient $T_{\rm ew}$,
jet power $L_{\rm j}$,
injected energy $E_{\rm j}/E_{\rm th}$, size of the domain
$l_{\rm domain}$, size of the domain in unit of the virial radius $l_{\rm domain}/R_{\rm vir}$, 
number of grid points $n_{\rm domain}$ and final time of the simulations $\tau_{\rm fin} = t/t_0$.}

\begin{center}
\begin{tabular}{ccccccccc}  \hline \hline
$M$ & $T_{\rm ew}$ (keV) & $L_{\rm j}$ (erg s$^{-1}$) & $E_{\rm j}/E_{\rm th}$ & $l_{\rm domain}$ (kpc) & 
$l_{\rm domain}/R_{\rm vir}$ & $n_{\rm domain}$ & $\tau_{\rm fin}$ \\

\hline

 105  &  2  &  $10^{46}$ & 0.068  & 819.2 & 0.55 & 2048 & 3.8\\
\hline

 150  &  1    &  $10^{46}$ & 0.39  & 819.2 & 0.77 & 2048  & 3.9 \\
\hline

 211 &   0.5    &  $10^{46}$ & 2.2 & 744.5 & 1 & 1860 & 2\\
\hline

 180 &  2  & $5\times10^{46}$ & 0.34 & 819.2 & 0.55 & 2048 & 2\\
\hline

 255 &  1  & $5\times10^{46}$ & 1.9 & 819.2 & 0.77 & 2048 & 1.95\\
\hline

 361 & 0.5 & $5\times10^{46}$ & 10.9 & 744.5 & 1 & 1860 & 0.85\\
\hline

\end{tabular}
\end{center}
\label{table:cases}
\end{table*}

We perform a first set of six simulations for three different X-ray temperatures ($T_{\rm ew}=0.5$, $1$, $2$ keV),
which correspond in our baseline model to virial masses $M_{\rm vir} = 2.3\times10^{13}$, $6.69\times10^{13}$, 
$1.89\times10^{14}$ $M_\odot$ 
respectively, and with two different jet kinetic powers ($L_{\rm j} = 10^{46}$, $5\times 10^{46}$ erg s$^{-1}$).
The masses of the clusters used as initial conditions for our simulations are different from
the virial masses of T03; we used T03 simulations only for building our baseline model.
In order to evaluate the effects on the large scale structures of an identical source of 
energy, the jets of equal power injected in the three groups/clusters have the same characteristic parameters
(i.e. density, speed and radius) and the two different powers are determined by their jet speed.
Since in our model the central density of the ambient medium is a constant at
all scales we choose then a constant $\nu=0.001$ for all these
simulations. Our simulated cases
are then characterized by their respective Mach number.
In all of our simulations we choose a square computational domain with a resolution of
$250$ points for $100$ kpc: the size of the computational domain $l_{\rm domain}$ is determined by the
minimum between the virial radius of the structure and the maximum length scale that we
decided to simulate ($819.2$ kpc or equivalently $2048$ points). The details about the computational
domain can be found in Table \ref{table:cases}.
In this set of simulations the jet is injected into the computational domain
for a time $t_{\rm life}=0.1 \; t_0 \simeq 2 \times 10^7$ years and is
then switched off imposing
reflecting boundary conditions over the whole boundary of injection ($z=0$).
In this way the total energy injected into the computational domain is of the
order of $E_{\rm j}\simeq 1.2\times10^{61}$ or $E_{\rm j}\simeq 6\times 10^{61}$ erg depending
on the jet power considered and taking into account the energy injected by both the jet and the
counterjet. This energy has to compete with the huge thermal energy of
the group or cluster $E_{\rm th}$ which for our baseline model is given by
\begin{equation}
E_{\rm th}=3.1 \times 10^{61} \left( \frac{T_{\rm ew}}{\rm keV} \right)^{5/2} 
\; {\rm erg}
\label{eq:scalen}
\end{equation}
After the jets have ceased their activity at $t/t_0=0.1$ the simulations are followed until 
the expanding bow shock determined by the energy injection covers at least $80\%$ of the 
computational domain.
In Table \ref{table:cases} we show the parameter $M$ for the
simulations that we have performed giving also the values of the ambient temperature
$T_{\rm ew}$, the energy injected into the domain $E_{\rm j}$ in units of the 
thermal energy $E_{\rm th}$ of the cluster, the size of the domain
$l_{\rm domain}$, the ratio between the size of the domain and the virial radius $R_{\rm vir}$, 
the number of grid points $n_{\rm domain}$ used for every of the two dimensions and the final
simulated time given in units of $t_0$.  

In order to understand how much the results are sensible to the jet parameters considered and to the resolution used, 
we also add three test simulations: using as initial conditions an unperturbed atmosphere with a mean temperature
$T_{\rm ew} = 0.5$ keV, we first repeat the case with $L_{\rm j} = 10^{46}$ erg s$^{-1}$ and $M=211$ with half the
spatial resolution. Then we
consider a jet with a kinetic power $L_{\rm j} = 10^{46}$ erg s$^{-1}$ injected for 
$t_{\rm life} = 0.1 t_0 \simeq 2 \times 10^7$ years as before but with a different Mach number and density ratio. 
Finally we simulate a jet with a lower kinetic power (lower density) injected for a longer time, taking the total 
injected energy equal to $E_{\rm j}\simeq 1.2\times10^{61}$ erg as before. 
In Table \ref{table:testcases} we show the parameters of these three simulations.

\begin{table}
\caption{Parameters of the three test simulations performed: Mach number $M$, 
density ratio $\nu = \rho_{\rm j}/\rho_0$, kinetic jet power $L_{\rm j}$,
active phase $t_{\rm life}$, number of grid points $n_{\rm domain}$ and final time 
of the simulations $\tau_{\rm fin} = t/t_0$. All these simulations have as initial
condition an atmosphere with $T_{\rm ew} = 0.5$ keV.}

\begin{center}
\begin{tabular}{cccccc}  \hline \hline
 $M$ & $\nu$ & $L_{\rm j}$ (erg s$^{-1}$) & $t_{\rm life}$ ($10^7$ yr) & $n_{\rm domain}$ & $\tau_{\rm fin}$ \\

\hline

211 &   0.001  &      $10^{46}$     & 2 & 930 & 3 \\
\hline

361  &  0.0002 &      $10^{46}$     & 2 & 1860 & 3 \\
\hline

211  &  0.0005 &  $5\times 10^{45}$ & 4 & 1860 & 2 \\
\hline

\end{tabular}
\end{center}
\label{table:testcases}
\end{table}

\section{The energy balance}
\label{sec:ene}

Firstly, we need to understand how the extragalactic jets interact with the 
surrounding medium and how the energy injected by the jet couples
with it.
Theoretical models (see Begelman and Cioffi 1989) and numerical simulations 
(see Norman et al. 1982, Massaglia et al. 1996) have analyzed in detail
the structure of the interaction between an underdense supersonic jet and the 
denser ambient: the jet is slowed down by a strong terminal shock (Mach disk) 
where its material is thermalized and inflates an expanding cocoon that compresses
and heats the ambient medium. 
The compressed ambient medium forms a shell surrounding the cocoon which
is separated from the shell itself by a contact discontinuity.
During this phase the cocoon and the shell are usually strongly overpressured with respect to 
the outer pressure (i.e. the initial ambient energy enclosed inside the 
shell's outer radius is negligible compared to the injected energy): 
the shell of compressed ambient material expands therefore with a supersonic speed
driving a strong shock into the surrounding medium where the expansion work done by the cocoon
of jet material is irreversibly dissipated.
In this regime the temporal behavior of the expanding shock can be obtained qualitatively 
assuming spherical symmetry (that is a good approximation at least for light jets; see
Zanni et al. 2003 and Krause 2003) and the  following behaviors for the injected energy 
$E_{j}$ and the density $\rho$ stratification:
\begin{equation}
E_{j} = E_0\left(\frac{t}{t_0}\right)^d \qquad \rho = \rho_0\left(\frac{r}{r_0}\right)^k  
\label{eq:injected}
\end{equation}
Since there are no characteristic lengths defined,
it is possible to derive the temporal behavior of the expanding shockwave only from
dimensional arguments (see Sedov 1959): we can write the shock $R_{\rm s}$ radius as
\begin{equation}
R_{\rm s} \propto \left(\frac{E_0 r_0^k}{\rho_0 t_0^d}\right)^\frac{1}{k+5}
t^\frac{d+2}{k+5} \qquad .
\label{eq:shradius}
\end{equation}
Assuming a polytropic relation for the ambient pressure $P \propto \rho^{\gamma}$ we can
derive the following behaviors for the shock speed $V_{\rm s}$ and Mach number $M_{\rm s}$:
\begin{equation}
V_{\rm s} \propto t^\frac{d-k-3}{k+5} \qquad \rm{and} \qquad 
M_{\rm s} \propto t^\frac{kd(1-\gamma)+2d-2\gamma k-6}{k+5} \qquad . 
\label{eq:mach}
\end{equation} 
It is easy to show that the Mach number decreases with time if the rate at which the energy 
is injected ($d$) is lower than the rate at which the background energy enclosed in the shock radius
($\propto \int_0^{R_{\rm s}} P r^2 \;dr$) grows: this means that as the shock expands the ambient energy
enclosed inside the shock volume becomes comparable to the injected energy and so 
the cocoon slowly reaches a pressure equilibrium  with the surrounding atmosphere expanding 
almost at the local sound speed.
The contact discontinuity, which in the overpressured phase expands at the
same rate of the outer shell radius (see Zanni et al. 2003), is subject to Rayleigh-Taylor
instabilities if its deceleration does not exceed the local gravity acceleration and 
to Kelvin-Helmholtz instabilities: these effects give rise to mixing between the jet material
which forms the cocoon and the shell  which surrounds it. Obviously, the entrainment with
the high entropy material of the cocoon is another source of irreversible (non--adiabatic) heating.
  
\begin{figure*}[ht]
\centering
\resizebox{17cm}{!}{\includegraphics{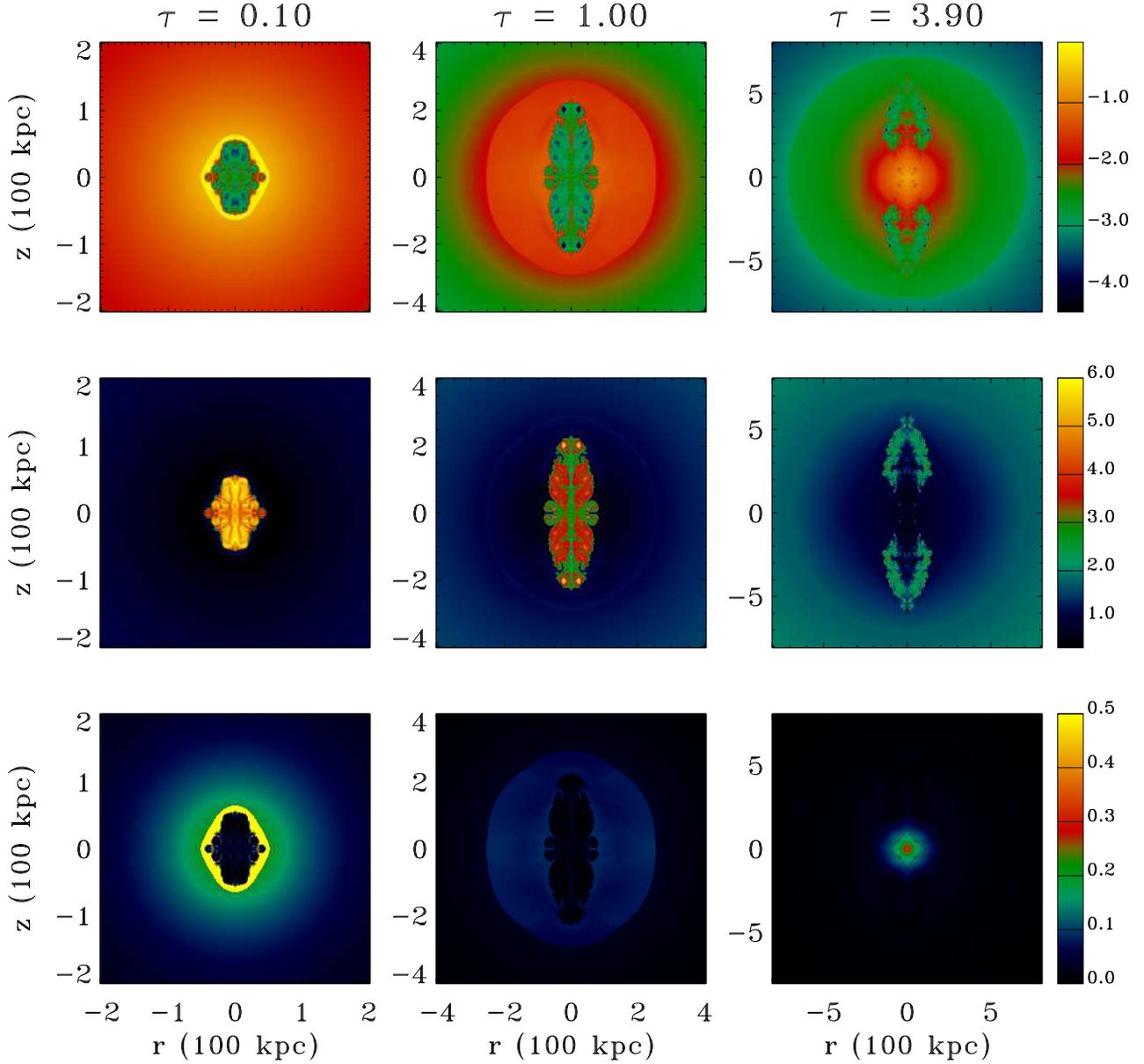}}
\caption{Time evolution of the $M=150$ case. {\it First row}: density maps in logarithmic
scale. {\it Second row}: specific entropy $S=T/n^{2/3}$ (logarithmic scale) which shows clearly the contribution 
of the entrainment by the high entropy material of the jet. {\it Third row}: entropy per unit volume 
($n \log(T/n^{2/3})$), which shows the contribution of the bow shock propagating into the medium. 
These quantities are shown in the columns at different times being $\tau=0.1$ the time at which
the jet is switched off and $\tau=3.9$ the final simulated time.
Notice the different spatial scale represented in the three columns.
The density and temperature are given in units of their values in the center of the undisturbed medium.  
}
\label{fig:figtot}
\end{figure*}

As described in Reynolds et al. (2002) once the jet has ceased its activity and is 
turned ``off'' it loses its collimation and subsequently collapses and disappears. 
The lobe of shocked jet material now rises into the heated atmosphere due to its
inertia and to its high entropy that makes it buoyantly unstable; the lobe evolves
into a rising ``plume'' partly entraining the ambient medium contained in the shell.
The bow shock, no more supported by the jet thrust, slows down and travels
across the ambient as a shock wave or as a sound wave, depending on its energy content
compared with the energy of the unperturbed medium. 

These subsequent stages of evolution are represented in Fig. \ref{fig:figtot} where 
we show maps of some interesting quantities at different times as they emerge from our simulation with $M=150$: 
in the first row we represent density maps in logarithmic
scale, in the second row we represent maps of entropy per particle ($S=T/n^{2/3}$) while, in the last row,
maps of entropy per unit volume ($S=n\log(T/n^{2/3})$) are shown. The first column shows maps of these quantities
at the end of the active phase ($\tau = 0.1$) while the other two columns show the evolution of the cocoon
once the jet has ceased its activity until the final simulated time ($\tau = 3.9$). Notice the different
spatial scale of the three columns. 
In the first row (density maps) we can visualize the different stages of evolution described above: during the active
phase (left panel) the jets push against the surrounding medium that is compressed and heated in a thin shell; in this
phase the injected energy is far greater than the background energy and so the shock driven by the expanding lobe is
very strong. Once the jets are turned off (central panel) the bow shock loses some of its strength but still expands as a mild
shock wave since its energy is still greater than the background: the post-shock material is beginning to expand no more
pushed by the expanding lobe. The lobes are stretching out due to their inertia and they
are entraining the surrounding shocked ambient gas (even if the mixing effect was present yet in the active phase 
due to KH instabilities). In the advanced phases of evolution the lobes are evolving as rising plumes of low density and high
entropy (see the second row) while the bow shock has lost quite completely its strength and so it mildly compresses the surrounding medium: in
fact, at this spatial scale, the injected energy, that as we will see after is quite completely transferred to the ambient 
medium, becomes comparable to the background one (see the $E_{\rm j}/E_{\rm th} \sim 0.4$ ratio in Table \ref{table:cases}).   
The shocked ambient material begins to recover a
new hydrostatic equilibrium displacing the rising lobes as can be clearly seen in the rightmost panel of the third row too
where the new core of higher entropy is clearly visible. 

In order to understand how much energy injected by the jet is transferred to the
ambient and in which form, we have defined the following integral quantities
\begin{equation}
\label{eq:energies}
\begin{array}{ccl}
\vspace{0.2cm}
E_{\rm int}(t) & = & \int \mathcal{T}\;\frac{p}{\gamma-1}\;dV \\
\vspace{0.2cm}
E_{\rm kin}(t) & = & \int \mathcal{T}\;\frac{1}{2}\rho v^2 \;dV \\
\vspace{0.2cm}
E_{\rm pot}(t) & = & \int \mathcal{T}\;\rho\phi \;dV \\
\vspace{0.2cm}
E_{\rm tot}(t) & = & E_{\rm int}(t) + E_{\rm kin}(t) + E_{\rm pot}(t) 
\end{array}
\end{equation}
which represent respectively the internal, kinetic, gravitational potential and total energy of the ambient medium;
in order to distinguish the ambient medium from the jet material, the integrals are performed weighing the energies 
per unit volume with the tracer $\mathcal{T}$ defined as the sum of the four tracers $\mathcal{T}_i$ given in Sect. \ref{sec:sim}
and therefore marking the entire cluster medium. The integrals are computed over the entire computational domain.   
Since the jet can transfer its energy to the ambient medium by means of the expansion work, done by
the cocoon of thermalized jet material, and due to the mixing with its high entropy material at the contact 
discontinuity, it is also important to quantify the energy transferred to the ambient by these two processes separately.
In the following discussion we will therefore indicate as ``compressed'' ambient medium
the ambient material which forms the shell surrounding the cocoon: this medium has been compressed both 
adiabatically and irreversibly (at the shock front) by the expansion work done by the cocoon.
On the other hand, we will indicate as the ``entrained'' ambient medium the shell ambient material which 
is subsequently entrained by the cocoon. 
As noticed before in Reynolds et al. (2002) the ambient medium which is entrained by the cocoon can be distinguished 
from the compressed ambient medium since its specific entropy $S=T/n^{2/3}$, at least in the first phases of evolution,
is much higher. 
Actually, the entropy per particle of the injected material is $10^5$ times higher than the entropy of the
center of the unperturbed cluster medium, considering a jet in pressure equilibrium with the cluster core and 
with a density ratio $\nu=0.001$ as in our main simulations. The entropy of the cocoon material is even higher since
it is formed by the jet material which has passed through the terminal shock. On the other hand the bow shocks 
which are found in our simulations can not raise the entropy per particle of the cluster core to values greater 
than $\sim 90$ (in units of the central value) which is the value of the entropy per particle of the unperturbed cluster 
at $r=R_{\rm vir}$. We therefore assume that the ambient material contained in the central part of the cluster 
which has an entropy per particle $S>S_0 = 90$ has been entrained by the cocoon. This separation is not
reliable at later times: when the bow shock reaches the outskirts of the cluster it raises the entropy to values
$S>S_0$ while numerical diffusion reduces the entropy of part of the cocoon below this threshold.
We therefore perform this analysis only for times $t/t_0 < 0.4$ taking $S_0 = 90$ as entropy threshold 
to distinguish between compressed and entrained ambient material. 

We can then define the total energy of the entrained (compressed) ambient material as:
\begin{equation}
E_{\rm tot, \;entr\;(comp)} = \int_{\rm E\; (C)} \mathcal{T} \;(\frac{p}{\gamma-1}+\frac{1}{2}\rho v^2+
\;\rho\phi) \; dV  
\end{equation}
where E and C are the volumes in which $S>S_0$ and $S<S_0$ respectively. 
Clearly, $E_{\rm tot,\; entr} + E_{\rm tot,\; comp} =E_{\rm tot}$.
 
\begin{figure}[ht]
\begin{center}
\resizebox{\hsize}{!}{\includegraphics{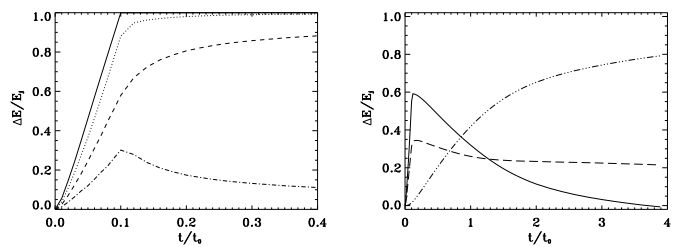}}
\caption{Temporal evolution of the variation of the ambient energy. 
({\it Left panel}) Temporal behavior of the total ambient energy at the beginning of the cocoon evolution
($t/t_0 < 0.4$). We show plots of the total injected energy ({\it solid line}), total ambient energy 
$E_{\rm tot}$ ({\it dotted line}),
total energy acquired by the shocked compressed material $E_{\rm tot,\; comp}$ ({\it dashed line}) 
and by the entrained material $E_{\rm tot,\; entr}$ ({\it dot--dashed line}).  
({\it Right panel}) Plotted here are the ambient
internal energy $E_{\rm int}$ ({\it solid line}), the ambient kinetic energy $E_{\rm kin}$ ({\it long--dashed line})
and the ambient gravitational potential energy $E_{\rm pot}$ ({\it triple--dot--dashed line}). 
Values of the energy are given in units of the total injected energy while time is given in units of $t_0$.
}
\label{fig:energyplot}
\end{center}
\end{figure}

We have plotted in Fig. \ref{fig:energyplot} the fractional difference between these
ambient energies at time $t$ and their initial values versus time. 
The values plotted are normalized with the value of the total injected energy. 

In the left panel of Fig. \ref{fig:energyplot} we plotted the temporal behavior of the total injected energy 
({\it solid line}) and the total energy transferred to the ambient medium $E_{\rm tot}$ ({\it dotted line}) 
in the initial stages of evolution ($t/t_0 < 0.4$): it is possible to see that in the active phase the jet can 
transfer a high fraction of its energy to the surroundings ($\sim 88\%$ at $t/t_0 = 0.1$). 
Plotted are also the increase of energy of the compressed ambient medium $E_{\rm tot,\; comp}$ ({\it dashed line}), 
which corresponds to the expansion work done by the cocoon, and the energy of the entrained ambient material 
$E_{\rm tot,\; entr}$ ({\it dot--dashed line}).
As we can see at the end of the active phase ($t/t_0 = 0.1$) the work done by the expanding cocoon 
corresponds approximatively to $60\%$ of the total injected energy. This fraction agrees with the following 
simple analytical estimate (given also by Ruszkowski et al. 2003 in the case of an expanding bubble in 
pressure equilibrium with the surroundings). We assume the following energy balance for the cocoon:
\begin{equation}
dE_{\rm j} = \frac{d(pV)}{\gamma -1}+pdV
\end{equation} 
where $E_{\rm j}$ is the energy injected in the cocoon, $p$ is its average pressure and $V$ is its volume and
so $d(pV)/(\gamma-1)$ is the variation of its internal energy and $pdV$ is the expansion work done at the contact
discontinuity. Assuming also the spherical self--similar behavior given by Eqs. (\ref{eq:injected}) and 
(\ref{eq:shradius}) for the injected energy, the atmosphere profile and the cocoon radius, we obtain the following
efficiency $\epsilon$ (ratio between the expansion work and injected energy):
\begin{equation}
\epsilon = \frac{pdV}{dE_{\rm j}}= \frac{(\gamma-1)(3d+6)}{\gamma(3d+6)+dk+2d-6}
\end{equation} 
which for $d=1$ (constant injection power) and $k=-2$ (isothermal sphere profile) gives $\epsilon = 66\%$.

At $t/t_0 = 0.1$ the energy transferred to the ambient medium by means of the mixing with the cocoon high entropy
material corresponds then to the remaining $\sim 28\%$ of the injected energy but it is possible to see that, just
after the jet stopped its activity, the entrained material tends to decrease its (mostly thermal) energy 
due to the expansion of the cocoon which is no more fueled by the jet. On the other hand, the ambient medium 
compressed by the still expanding cocoon increases its energy up to values $\sim 90\% E_{\rm j}$.     
As pointed out before, it is difficult to distinguish between the compressed and the entrained ambient 
medium on longer timescales, but based on these considerations, we can argue that at later times most of 
the injected energy is retained by the compressed ambient medium and only a small fraction by the rising lobes
and by the entrained ambient material.

Looking at the right panel of Fig. \ref{fig:energyplot} we can finally see in which form the energy is transferred
to the ambient medium on longer timescales. In this plot we make no distinction 
between compressed and entrained material. 
In the active phase of the jet ($t/t_0<0.1$) the ambient is compressed, 
heated by the expanding shock wave and partly entrained by the cocoon, thus increasing its internal $E_{\rm int}$ 
({\it solid line}), kinetic $E_{\rm kin}$ ({\it long--dashed line}) and gravitational potential energy $E_{\rm pot}$ 
({\it triple--dot--dashed line}) since its material is pushed higher in the potential well.
In the passive phase the ambient material, no more pushed by the internal cocoon, slows down, decreasing
its kinetic energy. Even if the bow shock is still expanding and heating the ambient, behind it the shocked
material can expand and decrease its internal energy trying to recover the hydrostatic equilibrium. Only the
gravitational energy is still growing since the shock still pushes the material higher in the potential well
while the expanding shocked material tries to recover the gravitational equilibrium on a higher adiabat thus
increasing its potential energy (it is less dense than it was in the initial equilibrium).

Since not all the energy that is transferred to the intracluster medium is dissipated, what it is more important 
to determine is the energy that is irreversibly transferred into 
the ambient medium or rather the heating that determines the increase of the entropy of the ICM.
As we have pointed out in this section the entropy of the ambient medium can be raised in two ways: shocked by the
blastwave pushed by the expanding cocoon or through the interaction with the high entropy material of the rising lobes.
In the following Section we will quantify the energy transfer associated with these two phenomena.   

\section{Entropy evolution}
\label{sec:entropy}

\begin{figure}[t]
\begin{center}
\resizebox{\hsize}{!}{\includegraphics{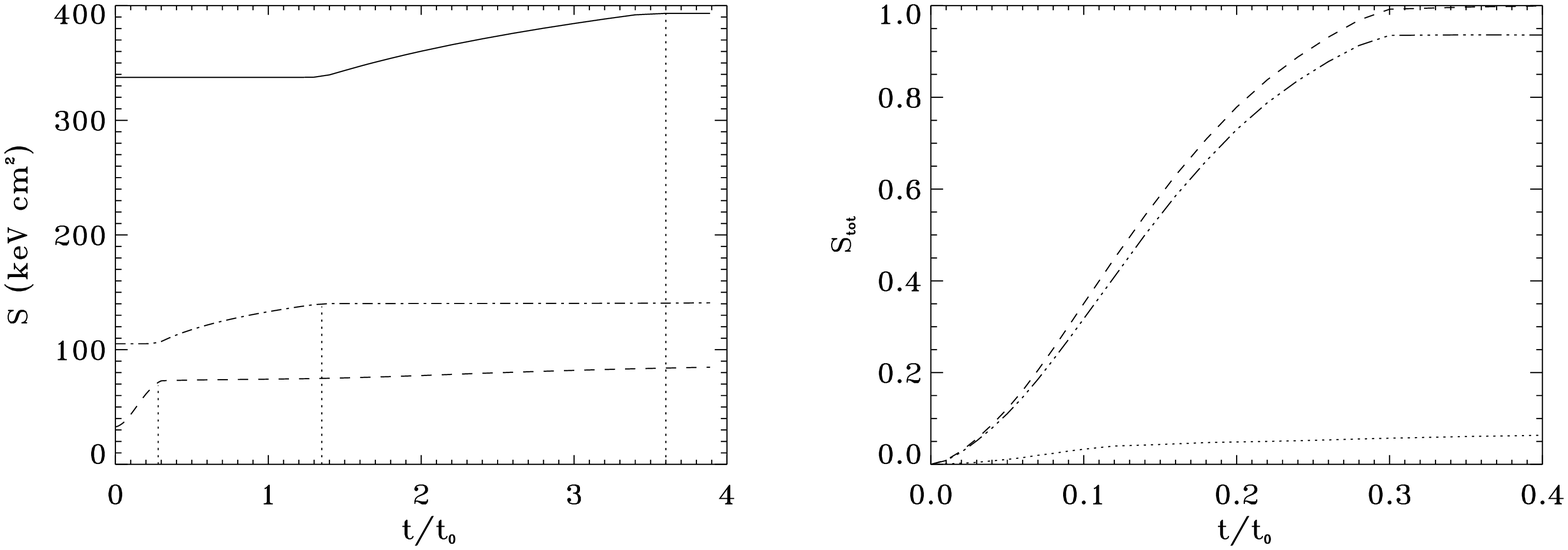}}
\caption{
({\it Left panel}) Temporal evolution of the average entropy per particle S=$T/n^{2/3}$ of the ambient material
marked by the innermost three tracers: $\mathcal{T}_1$ ($r/r_{\rm s} \in \; [0,1]$, {\it dashed line}), $\mathcal{T}_2$ ($r/r_{\rm s} 
\in \; ]1,3]$, {\it dot-dashed line}) and $\mathcal{T}_3$ ($r/r_{\rm s} \in \; ]3,6]$, {\it solid line}). 
({\it Right panel}) Temporal evolution of the total entropy increase $S_{\rm tot}$ (Eq. (\ref{eq:Stot})) of the innermost
tracer $\mathcal{T}_1$: plotted are the total entropy increase of the ambient medium ({\it dashed line}), the entropy increase
of the shocked material ({\it triple--dot--dashed line}) and the entropy increase of the entrained material 
({\it dotted line}). 
Time is given in units of $t_0$.}
\label{fig:entrevol}
\end{center}
\end{figure}

We can understand how the ambient entropy has increased due to the interaction with the jets looking at the
second and third row of Fig. \ref{fig:figtot}: the maps in the second row 
(evolution of the entropy per particle $S=T/n^{2/3}$) show clearly how the entrainment with the high entropy material
of the jets can raise substantially the entropy of the mixed material while the entropy increase of the shocked gas is 
not even noticeable. But if we look at the entropy per unit volume (third row) we see clearly the expanding shock to be 
the dominant feature: this means
that the entrainment with the high entropy material of the lobes increases the entropy of the ambient to higher values
but involves only a little fraction of the mass of the ICM; on the other hand the expanding shock wave increases the ambient 
entropy to lower values involving a greater fraction of the ICM traveling across a big portion of the structure.
This behavior can be seen more quantitatively in the left panel of Fig. \ref{fig:entrevol} where we have plotted
the evolution of the mean entropy per particle of the first three tracers that follow the evolution of the ambient medium:
we can see that the entropy increase of the tracers is mostly associated with the passage of the bow shock since once the
shock has overtook the inner tracer this one evolves adiabatically while the outer one starts to increase its entropy due
to the shock passage. 
To evaluate the importance of entrainment in rising the entropy of the ambient medium 
we can define separately the variation of the total entropy of the entrained material, characterized by
an entropy per particle $S>S_0=90$, and of the compressed material, which has $S<S_0$. We remember that
this kind of analysis is not correct on longer timescales.  
We plotted in the right panel of Fig. \ref{fig:entrevol} the variation of the total entropy 
\begin{equation}
S_{\rm tot} = \int \mathcal{T}_1 \; n\log\left(\frac{T}{n^{2/3}}\right)\;dV
\label{eq:Stot}
\end{equation}
of the inner tracer at the beginning of the cocoon evolution ($t/t_0 <0.4$) ({\it dashed line}, values are normalized to the 
plateau reached by this curve).
Computing the integral of Eq. (\ref{eq:Stot}) over the volumes in which $T/n^{2/3}<S_0$ and
$T/n^{2/3}>S_0$, 
we plotted respectively also the variation of entropy of the ambient material compressed by the bow-shock 
({\it triple--dot--dashed line}) and the variation of entropy of the entrained material ({\it dotted line}): we can see that the
shocked material can account for a high fraction of the total increase of entropy ($\sim 94\%$) while the contribution
of the entrained medium is negligible ($\sim 6\%$).
 
\begin{figure}[ht]
\resizebox{\hsize}{!}{\includegraphics{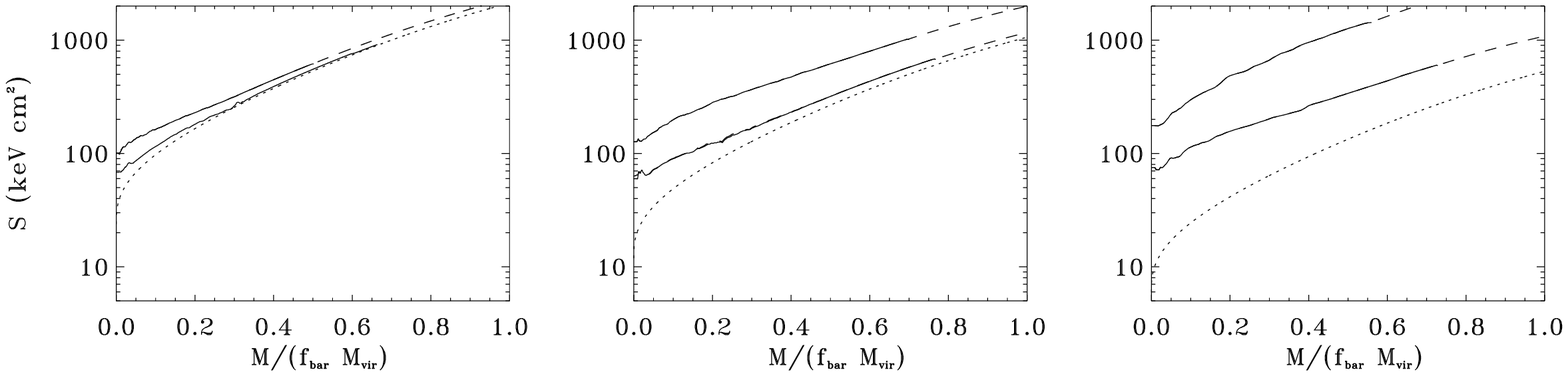}}
\caption{Entropy $S=T/n^{2/3}$ profiles as a function of Lagrangian (mass) coordinate at the beginning and at the end of 
the simulation. The panels refer to the simulations performed with the $T_{\rm ew}$ = 2 keV ({\it left panel}), 
$T_{\rm ew}$ = 1 keV ({\it central panel}) and $T_{\rm ew}$ = 0.5 keV ({\it right panel}) clusters. 
In each panel are shown the initial
entropy profile ({\it dotted line}), the profiles determined by the interaction with the jets characterized by a luminosity
$L_{\rm j}= 10^{46}$ erg s$^{-1}$ ({\it lower solid line}) and $L_{\rm j}= 5\times 10^{46}$ erg s$^{-1}$ 
({\it upper solid line}) and analytical extrapolations of the profile up to the virial radius ({\it dashed line}).
The mass coordinate is given in units of the total gas mass content of the clusters $f_{\rm bar} M_{\rm vir}$.}
\label{fig:entropyplot}
\end{figure}

On this basis we can argue that in our simulations the shock heating is the dominant effect determining 
the thermodynamical evolution of the ICM.
Since the propagating shock, mostly after the jet has been turned off, assumes a spherical shape, we can define
an entropy profile taking averages over spherical shells of given mass, building an entropy profile as a function
of the mass enclosed inside a given radius. 
In Fig. \ref{fig:entropyplot} we plot the entropy profile associated with 
the passage of the shock wave as a function of the enclosed mass for the six cases 
of Table \ref{table:cases}: the three panels represent the three clusters taken into account ($T_{\rm ew}$ = 2, 1, 0.5 keV
from left to right) while the solid lines in each panel represent the entropy profiles determined by
the two jet powers considered ($10^{46}$ and $5\times 10^{46}$ erg s$^{-1}$ for the lower and upper curves
respectively). The unperturbed initial entropy profiles are shown in each panel with a dotted line. 

In the cases in which the ambient energy enclosed inside the shock radius  becomes comparable 
to the almost constant energy associated with the shock wave, 
the shock loses its strength and evolves slowly as a sound wave, not increasing the ambient entropy from a 
certain radius (mass) onward. From this point onward the entropy profile nearly matches that of the unperturbed
medium (see for example the lower solid lines in the left and middle panel in Fig. \ref{fig:entropyplot}). 
For the cocoons that remain overpressured for the entire evolution inside the cluster or
group, the expanding shock never loses its strength and the entropy of
the atmosphere is increased everywhere.
In some of our simulated cases, we cannot follow the evolution of the
blastwave up to the the virial radius, since the virial radius is not
contained inside the computational domain of some simulations, or it is not
reached at the final time of the simulation.
In Fig. \ref{fig:entropyplot} the entropy profiles derived directly from the
numerical simulations  ({\it solid lines}) stop at the farthest coordinate values reached.  
In this situations,
we build an extrapolation of the entropy profile from the maximum radius reached in the simulation up to the virial
radius in the following way: we assume that in these final stages of evolution the energy $E$ associated with the blast
wave is a constant; we assume that a constant fraction of this energy goes into thermal energy (pressure) and that
this pressure ($>$ external pressure) drives the expansion of the shock. We can then write the following
expressions for the pressure $P_{\rm s}$ driving the shock and the radius $r_{\rm s}$ of the shock (see 
Appendix A in Zanni et al. 2003):
\[
P_{\rm s} \propto \frac{E}{r_{\rm s}^3} \qquad \qquad \qquad \frac{dr_{\rm s}}{dt} \propto \sqrt{\frac{P_{\rm s}}{\rho(r_{\rm s})}}
\]
where $\rho(r_{\rm s})$ is the ambient density evaluated at the shock radius $r_{\rm s}$.
This set of equations gives the following solution:
\begin{equation}
t-t_0=K\int_{r_0}^{r_{\rm s}}\rho(r)^{\frac{1}{2}}r^{\frac{3}{2}}dr
\label{eq:shockrad}
\end{equation}   
where $r_0$ and $t_0$ are the shock radius and the time at which we begin to extrapolate the entropy
profile while $K$ is a constant that is determined matching the solution for the averaged shock radius
coming from the simulations with the solution of Eq. (\ref{eq:shockrad}). Once we have determined the shock radius as a function of time
we can derive its speed and its Mach number with respect to the ambient sound speed. From the Mach number
of the shock we can finally derive the entropy jump associated with the expanding wave. These extrapolated
profiles are represented in Fig. \ref{fig:entropyplot} with dashed lines.

\begin{figure}[ht]
\resizebox{\hsize}{!}{\includegraphics{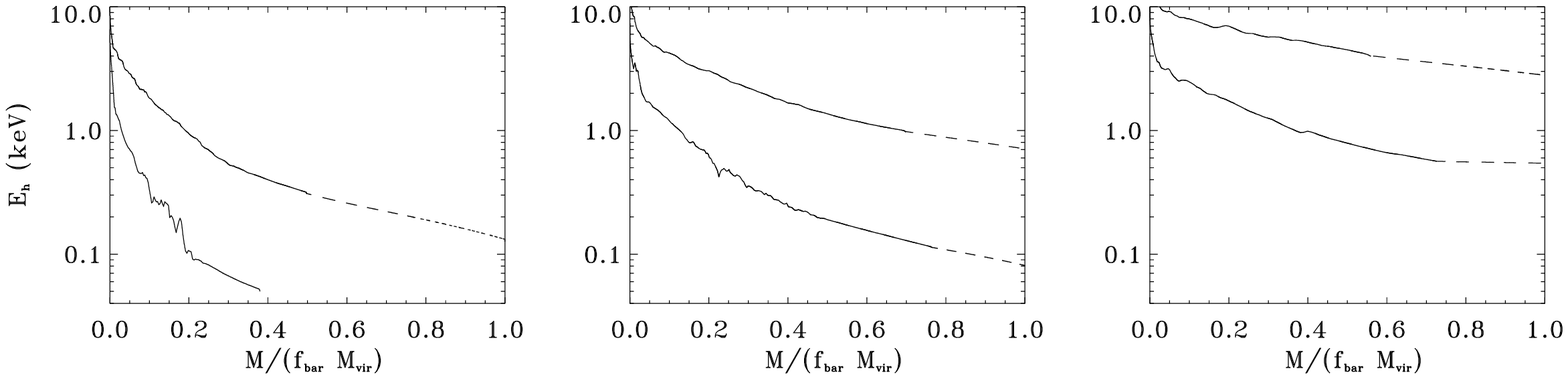}}
\caption{Plot of the heating energy per particle $E_{\rm h}$ needed to obtain the entropy profiles of Fig. \ref{fig:entropyplot}
as a function of mass coordinate. The panels refer to the $T_{\rm ew}$ = 2 keV ({\it left panel}), 
$T_{\rm ew}$ = 1 keV ({\it central panel}) and $T_{\rm ew}$ = 0.5 keV ({\it right panel}) clusters. 
In each panel we plot the heating energy per particle given by the $L_{\rm j}= 10^{46}$ erg s$^{-1}$ 
({\it lower solid line}) and $L_{\rm j}= 5 \times 10^{46}$ erg s$^{-1}$  jets ({\it upper solid line})
and analytical approximations up to the virial radius ({\it dashed line}) obtained from the extrapolated entropy
profiles.
The mass coordinate is given in units of the total gas mass content of the clusters $f_{\rm bar} M_{\rm vir}$.  }
\label{fig:tdsplot}
\end{figure}

\begin{figure}[ht]
\resizebox{\hsize}{!}{\includegraphics{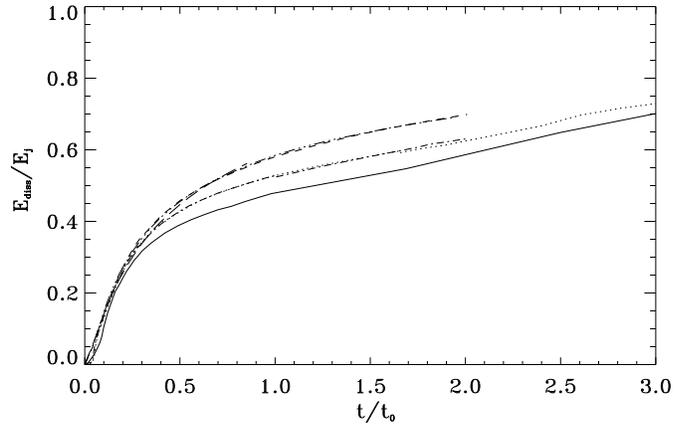}}
\caption{Plot of the total heating energy dissipated by the jet in units of the total injected energy as a function
of time. The curves refer to the cases characterized by $M=105$ ({\it solid line}), $M=150$ ({\it dotted line}), $M=211$ 
({\it dashed line}), $M=180$ ({\it dot-dashed line}), $M=255$ ({\it double dot-dashed line}) and $M=361$ ({\it long
dashed line}). It is important to notice that the lines are divided in three groups characterized by different ratios
$E_{\rm j}/E_{\rm th}$ between the total injected energy and the thermal energy of the cluster: $E_{\rm j}/E_{\rm th}$ =
10.9, 2.2, 1.9 (upper lines), $E_{\rm j}/E_{\rm th}$ = 0.39, 0.34 (middle lines) and $E_{\rm j}/E_{\rm th}$ = 0.068 
(lower line). See Table \ref{table:cases} for the correspondence between $E_{\rm j}/E_{\rm th}$ and Mach number. }
\label{fig:tdsevol}
\end{figure}

From these entropy profiles it is possible to define an ``equivalent heating'' assuming that these profiles
are obtained from an isochoric transformation, i.e. keeping constant the initial density profile of the
ambient and increasing only its temperature. This quantity gives an information on the amount of energy that
is irreversibly dissipated inside the cluster by the jet and goes into heat. This estimate is obviously different from the
real amount of injected energy that is converted into heat, since the thermodynamical transformation associated
with the propagation of the shock which is compressing the gas is not isochoric.
Anyway this is exactly the amount of energy gained by the gas particles in a non--adiabatic compression followed
by an adiabatic expansion to recover the initial density.  
In Fig. \ref{fig:tdsplot} we show a radial profile (in mass
coordinates) of this equivalent heating, 
showing the different amount of heating energy per particle $E_{\rm h}$ released
inside the clusters for our six cases:
in the three panels we show the results for the three clusters
considered ($T_{\rm ew}=$   2, 1, 0.5 keV from left to right),
and in each panel the solid lines represent the profiles obtained
with the two jet powers taken into account
($10^{46}$ and $5\times 10^{46}$ erg s$^{-1}$ for the lower and upper
curves respectively ). The dashed lines
represent the heating derived from the extrapolated entropy profiles.
In Fig. \ref{fig:tdsevol} we show the temporal behavior of the spatially integrated value 
of this quantity normalized to the total injected energy, showing that in all the six cases 
considered a high fraction (up to $\sim 75 \%$) of the injected energy 
is dissipated through irreversible processes and heats the cluster.
In Fig. \ref{fig:energyplot} we showed that almost all the injected energy is transferred to the ambient
medium: we find a lower ``equivalent heating'' since it accounts only for processes 
which can raise the entropy of the ICM and therefore neglects the energy which is transferred 
adiabatically.
The upper curves in Fig. \ref{fig:tdsevol} refer to the cases with a
$E_{\rm j}/E_{\rm th}$ ratio equal to 10.9, 1.9, 2.2,
the middle curves to the cases characterized by $E_{\rm j}/E_{\rm th}$
= 0.39, 0.34 while the lower solid line represents the $E_{\rm j}/E_{\rm th}$ = 0.068 case.
The jets characterized by a low $E_{\rm j}/E_{\rm th}$ ratio can also
efficiently dissipate their
energy, since they can be in an overpressured phase at the beginning of
their evolution. Observations of the diffuse X-ray thermal emission 
associated with the low power radio source NGC1052 (Kadler et al. 2004) 
could confirm this possibility.

\section{Hydrostatic equilibrium}
\label{sec:equilibrium}

\begin{figure}[ht]
\resizebox{\hsize}{!}{\includegraphics{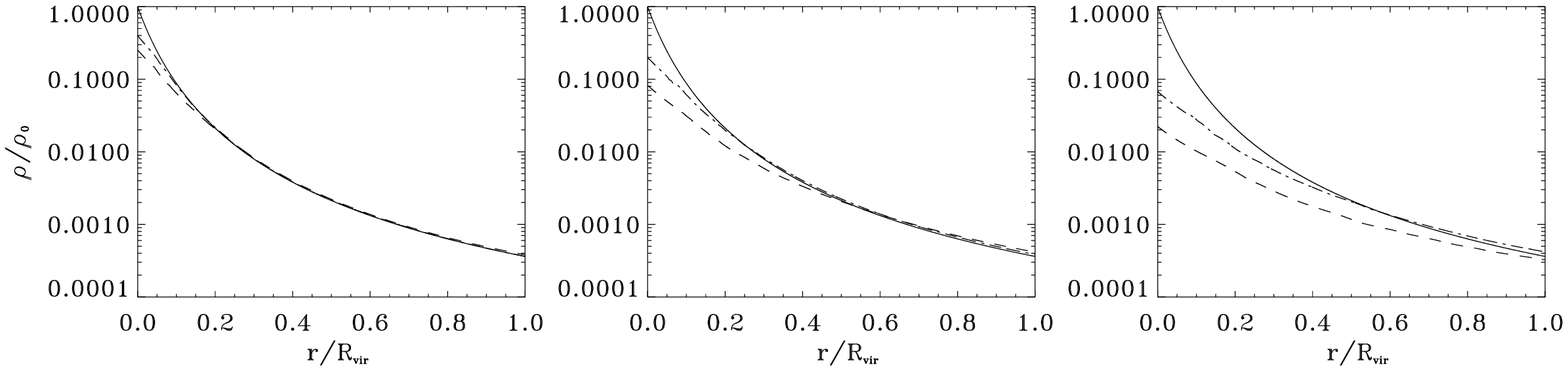}}
\caption{Radial density profiles of the clusters defined in the initial conditions (Eq. (\ref{eq:rhogas})) 
and obtained by calculating the hydrostatic equilibrium determined by the entropy profiles of Fig. \ref{fig:entropyplot}.
The panels refer to the $T_{\rm ew}$ = 2 keV ({\it left panel}), $T_{\rm ew}$ = 1 keV ({\it central panel}) and 
$T_{\rm ew}$ = 0.5 keV ({\it right panel}) clusters. Plotted are the profiles of our baseline model ({\it solid line}) and
the profiles determined by taking into account the energy dissipated by the $L_{\rm j}= 10^{46}$ erg s$^{-1}$ 
({\it dot-dashed line}) and $L_{\rm j}= 5 \times 10^{46}$ erg s$^{-1}$  jets ({\it dashed line}). The density is
given in units of the central density in the initial conditions while the radial coordinate is normalized with the
characteristic radius $r_{\rm s}$.}
\label{fig:densplot}
\end{figure}

Once we have calculated the entropy profiles as a function of the included mass, assuming
that the atmosphere will recover the hydrostatic equilibrium without additional dissipation,
we can determine the radial behavior of all the
thermodynamical quantities in the new hydrostatic equilibrium solving the following system of equations:
\begin{displaymath}
\left\{
\begin{array}{ccc}
\frac{\partial P}{\partial r} & = & \left(\frac{P}{S(M)}\right)^\frac{1}{\gamma} g \\
 & & \\
\frac{\partial M}{\partial r} & = & 4\pi r^2 \left(\frac{P}{S(M)}\right)^\frac{1}{\gamma} 
\end{array}
\right.
\end{displaymath}
where we used the relation $S(M)=P/\rho^\gamma$, where $S(M)$ is the averaged entropy profile as a function of
the included mass as calculated from our simulations. The gravity acceleration $g$ is derived 
from the potential of Eq. (\ref{eq:potdm}).
In order to solve this system of equations we need two boundary conditions being the first, obviously, 
$M(r=0) = 0$. The second boundary condition is given requiring that the thermal pressure found in correspondence
of the virial (gas) mass is the same given in the initial profile, assuming, in a sense, that the conditions
outside the cluster did not change. However, we found that the quantities that we calculated from these final
profiles did not depend strongly on the assumed boundary conditions: our outer boundary condition requires
only that the pressure outside the cluster is much lower than the central one. The central values of the interesting
thermodynamical quantities (density, pressure and temperature) depend only slightly on the boundary condition 
assumed, while the integral quantities such as the gas mass contained inside the virial radius, as noticed 
before by Voit et al. (2002), is more sensible.  
As an example we plot in Fig. \ref{fig:densplot} the density profiles of the new hydrostatic equilibrium for the six
simulations of Table \ref{table:cases}: in the three panels we show the results for the three clusters considered 
($T_{\rm ew}$ = 2, 1, 0.5 keV from left to right), the dot-dashed and the dashed lines represent the profiles obtained with the $10^{46}$ 
and $5\times10^{46}$ erg s$^{-1}$ jets respectively.

\begin{figure}[ht]
\resizebox{\hsize}{!}{\includegraphics{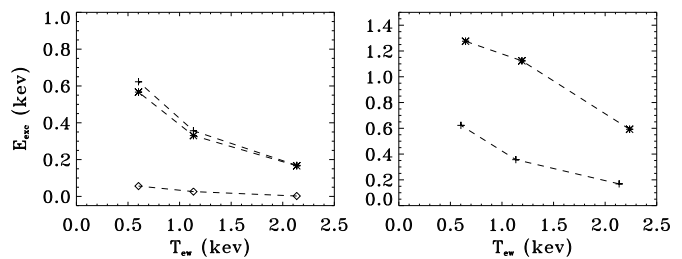}}
\caption{Plot of the difference between the average energy per particle of the final and initial hydrostatic equilibrium
as a function of the final emission-weighted temperature of the clusters. The average of the energy excess is calculated over
the particles contained inside the virial radius in the final hydrostatic conditions. ({\it Left panel}) Plot of the
average energy excess per particle determined by the $L_{\rm j} = 10^{46}$ erg s$^{-1}$ jets: plotted are the total energy
({\it crosses}), the gravitational potential energy ({\it asterisks}) and the internal energy ({\it diamonds}). 
({\it Right panel}) Plot of the total energy excess per particle determined by the $L_{\rm j}= 10^{46}$ erg s$^{-1}$ 
({\it crosses}) and the $L_{\rm j}= 5 \times 10^{46}$ erg s$^{-1}$ jets ({\it asterisks}).}
\label{fig:excess}
\end{figure}

Another way to give an estimate of the energy per particle that the jet has dissipated into the ambient medium is to 
evaluate the difference between the average energy per particle (thermal and potential) in the final and the initial
hydrostatic conditions.  Moreover, this estimate does not depend on the assumption of an isochoric transformation
on which the calculation of the ``equivalent heating'' is based.
In Fig. \ref{fig:excess} we plot in the left panel the excess energy $E_{\rm exc}$ of 
the particles contained inside the virial radius in the final state versus
the temperature of the cluster due to the $10^{46}$ erg s$^{-1}$ jets: the total energy is shown with the crosses while 
the asterisks and the diamonds correspond respectively to the potential and thermal excess energy. We can see that in 
the final state the greater amount of excess energy is in the form of potential energy since the heating of the
ICM has led to a huge expansion of the medium thus increasing its gravitational energy (the final state is less dense
then the initial one and the gravitational potential is the same). The expansion of the structure has obviously led
to a substantial adiabatic cooling and so the final thermal energy is not much greater than the initial one. In the right
panel of Fig. \ref{fig:excess} we plot the total excess energy as a function of the mean temperature of the clusters
as due to the $10^{46}$ (crosses) and $5\times10^{46}$ erg s$^{-1}$ (asterisks) jets. We can see that
the excess energy is lower than the energy per particle dissipated in the computational domain (Table \ref{table:tdiss}): 
this is due to the work done in the expansion to move the mass that now is located outside the virial radius and so this difference 
is more evident in the cases that have expanded more. 

\begin{figure}[t]
\resizebox{\hsize}{!}{\includegraphics{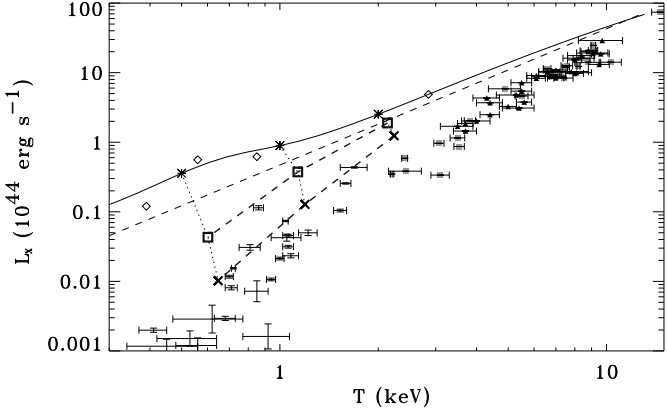}}
\caption{Bolometric luminosity-emission weighted temperature relation for our simulations 
characterized by $L_{\rm j}= 10^{46}$ erg s$^{-1}$ ({\it empty squares}) and $L_{\rm j}= 5 \times 10^{46}$ erg s$^{-1}$ 
({\it crosses}) in comparison with observational data ({\it pluses}: Helsdon \& Ponman 2000; {\it filled squares}: 
Arnaud \& Evrard 1999; {\it filled triangles} Markevitch 1999). 
Plotted are also the scaling of the bremsstrahlung luminosity ({\it dashed line}) and the X-ray luminosity with a 
correction for line emissivity (metallicity $0.3Z_\odot$) ({\it solid line}) of our initial model.
The bolometric luminosities of the three initial unperturbed clusters ({\it asterisks}) and of the four Tornatore
et al. (2003) simulations ({\it diamonds}) from which our initial conditions were derived are shown for comparison.
}
\label{fig:lumt}
\end{figure}

\begin{figure}[t]
\resizebox{\hsize}{!}{\includegraphics{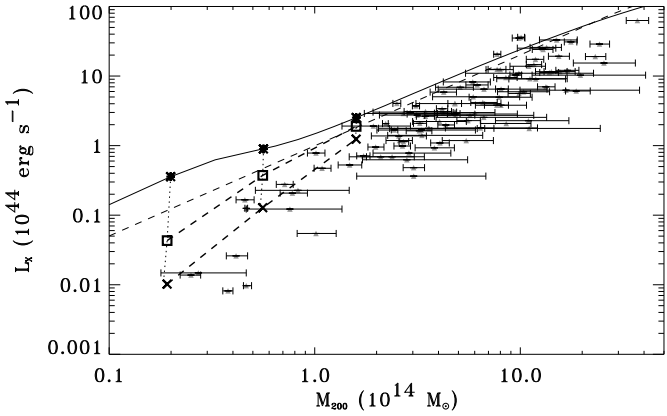}}
\caption{Comparison between the X--ray luminosity - $M_{200}$ relation obtained from our simulations with
$L_{\rm j}= 10^{46}$ erg s$^{-1}$ ({\it empty squares}) and $L_{\rm j}= 5 \times 10^{46}$ erg s$^{-1}$ 
({\it crosses}) and the observational data ({\it triangles}: Reiprich \& B\"ohringer 2002). 
Plotted are also the scaling of the bremsstrahlung luminosity ({\it dashed line}) and the X-ray luminosity with a 
correction for line emissivity (metallicity $0.3Z_\odot$) ({\it solid line}) of our initial model.
The bolometric luminosities of the three initial unperturbed clusters ({\it asterisks}) are also plotted.}
\label{fig:lumm}
\end{figure}

With these new profiles it is possible to determine how the observable properties of the clusters have 
changed consequently to the jet energy injection. 
We then computed the X-ray luminosity adding to the bremsstrahlung emissivity a correction for 
metal--line cooling assuming a plasma with $Z=0.3Z_\odot$ metallicity. In Fig. \ref{fig:lumt} we plotted
the results obtained for the X--ray luminosity - temperature relation while the comparison between the observed and 
the simulated $L_X - {\rm M}_{200}$ relation is shown in Fig. \ref{fig:lumm}. In both Figures we also plotted the
scaling of our initial model for the X--ray luminosity including line emission ({\it solid line}) and for 
bremsstrahlung luminosity only ({\it dashed line}): we recall that this last quantity satisfies the self--similar 
behaviors $L_{\rm br} \propto T_{\rm ew}^2$ and $L_{\rm br} \propto M_{200}^{4/3}$ respectively. 
It is possible to see that the luminosities obtained with the higher power $L_{\rm j}= 5 \times 10^{46}$ erg s$^{-1}$ 
jets are in good agreement with the observational data, mostly at the smaller simulated scales.
From Fig. \ref{fig:lumt} we can also notice that, as discussed before, the temperature of the final state is only slightly higher than 
the initial one and the lower X--ray luminosity is therefore mainly determined by the lower density of the final
state. 

\begin{figure}[ht]
\resizebox{\hsize}{!}{\includegraphics{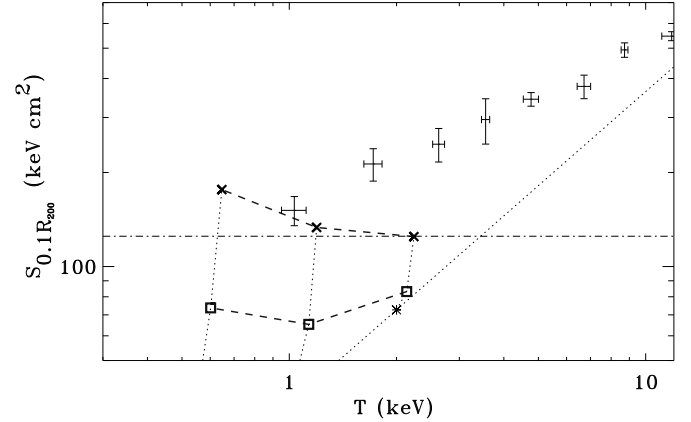}}
\caption{Entropy (at $0.1R_{200}$) - emission weighted temperature relation for our simulations
($L_{\rm j}= 10^{46}$ erg s$^{-1}$, {\it empty squares} and $L_{\rm j}= 5 \times 10^{46}$ erg s$^{-1}$,
{\it crosses}), in comparison with observational data ({\it pluses}: Ponman et al. 2003).
The observational data represent averages in different bins of temperature.
The dotted line represents the scaling of the initial model (Eq. (\ref{eq:scalrel})). 
The dotted-dashed line represents the constant floor 125 $(h/0.7)^{-1/3}$ keV cm$^2$ 
({\it crosses}) determined by Lloyd-Davies et al. (2000). }
\label{fig:entrt}
\end{figure}

On the other hand (see Fig. \ref{fig:lumm}), $M_{200}$ is poorly affected by the gas mass loss, since the
mass content of the cluster is mainly determined by the dark matter. 
While the observed $M_{200}$ data represent the gravitational mass contained inside $R_{200}$, 
we determined $M_{200}$ in our simulations adding up the dark matter content
and the gas mass, which is not self--gravitating. Due to the low gas mass fraction of these
structures this should not affect dramatically our results.
Finally, since neither the average temperature nor the total mass content of the clusters
are strongly affected by the energy injection,
the $T_{\rm ew} - M_{200}$ relation does not deviate much from the 
$T_{\rm ew} \propto M_{200}^{2/3}$ self--similar behavior, consistently with the observations (see Horner et al. 
1999 and Reiprich \& B\"ohringer 2002). 
In Fig. \ref{fig:entrt} we plot the relation between the entropy per particle estimated at $0.1 R_{200}$ 
and the mean temperature. As previously noticed for the X--ray luminosity relations, also the increase of entropy
is affected more by the lower density of the final state than by the slightly higher temperature.
In this plot is even more evident that a good agreement with the data is obtained only with the higher power  
energy injections at the smaller scales ($T_{\rm ew} < 1$ keV). This in not surprising: it has been argued 
by Roychowdhury et al. (2004) that an energy injection which increases linearly with the cluster mass is needed to
reproduce the entropy profiles observed by Ponman et al. (2003): on the other hand we are injecting the same 
amount of energy at all scales which does not seem to be enough for the more massive clusters.

\section{Tests on numerical resolution and jet parameters.}

We finally discuss the results obtained with the ``test'' cases
reported in Table \ref{table:testcases}.
We recall that these simulations have been performed in order to test
how much the results obtained with our six simulations depend on the
grid resolution and on the jet parameters assumed; we therefore took a reference 
case with an initial atmosphere with $T_{\rm ew} = 0.5$ keV
($M=211$, third line in Table \ref{table:cases}) and perform three additional runs.
The first one (first line in Table \ref{table:testcases}) is a case with
the same jet parameters (Mach number,
jet density, jet radius and duration of injection) of the reference
simulation but with half the grid resolution:
we don't find significant differences in the entropy profiles (and
therefore in derived results), showing
that the adopted resolution is high enough to obtain numerical
convergence.  On the other hand, it is well known that
shock capturing methods such as the PPM rapidly converge to a solution
in problems involving strong shocks as in our simulations.
The case represented in the second line of Table \ref{table:testcases} has the same jet
kinetic power and injection time of the reference simulation
but higher Mach number and lower density: also in this case, 
discrepancies between the resulting entropy profiles and the ones
described in the previous Sections are well within a few percent.
This indicates that the parameter controlling the jet dynamics is
the jet power, quite independently from other
parameters (Mach number, density and radius) which characterize the jet.  
This hypothesis is confirmed also by the fact that simulations characterized 
by similar ratios between the injected energy and the thermal energy of the unperturbed cluster 
$E_{\rm j}/E_{\rm th}$  but with different Mach numbers
and different ratios between the jet radius and the core radius $r_{\rm j}/r_{\rm s}$, have given similar results:
it has been yet noticed that similar efficiency curves (Fig. \ref{fig:tdsevol}) are obtained
from simulations with comparable $E_{\rm j}/E_{\rm th}$. Moreover Fig. \ref{fig:densplot} shows 
clearly that these simulations give similar final density profiles with approximatively
the same decrease in the central density (compare for example the dashed line in the central
panel with the dot--dashed line in the right panel which are obtained from two simulations characterized by
$E_{\rm j}/E_{\rm th}=$  1.9 and 2.2 respectively): this means
that if these cases are rescaled to the same initial conditions (same $E_{\rm th}$) they will 
give comparable observable quantities.

Different considerations can be done for the last case in Table
\ref{table:testcases} (third line) that is a jet with half
the kinetic power of the reference simulation (due to its lower
density) but injected for twice the time ($t/t_0 = 0.2$)
so as to have the same total injected energy. Since the strength of
the expanding shock is mainly determined by the
injected energy, it is easy to guess that in the central part of the
group the shocks driven by the lower power case will 
be weaker, while on a larger scale, since the energy injected by the
two jets is the same, the shocks will have more or less the
same strength. This is confirmed by Fig. \ref{fig:diff} where we plot
the entropy profiles given by the lower power ({\it solid line}) and
the higher power jet ({\it dashed line}) as a function of the mass coordinate:
\begin{figure}[ht]
\resizebox{\hsize}{!}{\includegraphics{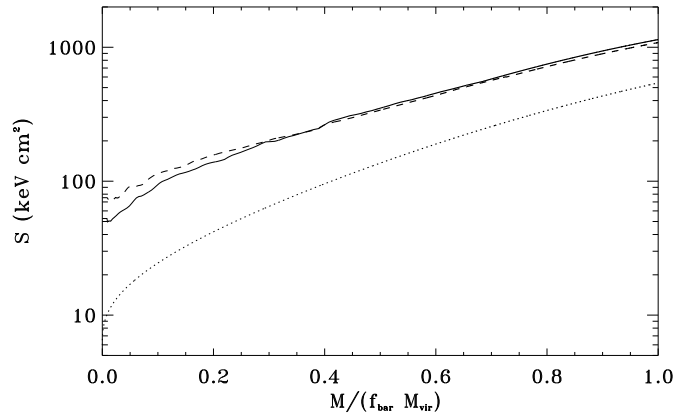}}
\caption{Entropy $S=T/n^{2/3}$ profiles as a function of Lagrangian (mass) coordinate. Plotted are 
the initial conditions ({\it dotted line}) and the results obtained with the $M=211$ simulation 
({\it dashed line}) which represents a $L_{\rm j}= 10^{46}$ erg s$^{-1}$ jet injected in a $T_{\rm ew}$ = 0.5
keV cluster for a time $t\sim 2\times 10^7$ years and with the case in the third line of Table
\ref{table:testcases} ({\it solid line})which is a jet with half the kinetic power (lower density) injected for twice the time.
The mass coordinate is given in units of the total gas mass content of the clusters $f_{\rm bar} M_{\rm vir}$.       
}
\label{fig:diff}
\end{figure}
while toward the central part the lower power jet determines a lower
entropy profile, on a larger scale the two profiles are absolutely
similar.  Also on the large scale the little discrepancy that can be seen
between the two profiles can be ascribed to the small difference
between the two total injected energies. The two
jets inject the same amount of kinetic energy but, having the same
internal energy flux, the lower kinetic power jet injects
twice the internal energy, giving a difference of a few percent on the
total injected energy. The difference found in the central part of
the profiles implies therefore a lower ($\sim 30\%$) ``equivalent
heating'' and leads obviously to a different new equilibrium, with a
higher central density and lower temperature. This produces a higher ($\sim 30\%$) 
X-Ray luminosity.

\section{Discussion and conclusions}
\label{sec:concl}

In the previous Sections we have shown how a (double-sided) powerful jet can dissipate efficiently
its bulk kinetic energy thus rising the entropy of the medium through which it propagates.  
This was done by analyzing by means of numerical hydrodynamical simulations
the interaction of two powerful supersonic jets with different powers ($L_{\rm j} = 10^{46}, 5\times10^{46}$ erg s$^{-1}$)
with the ICM of  three gravitationally heated clusters of galaxies characterized 
by different temperatures ($T_{\rm ew} = 0.5, 1, 2$ keV).
 
The mechanism through which these jets interact with the ambient medium can be summarized as follows: 
an underdense supersonic jet is slowed down by a terminal
shock (Mach disk) where its kinetic energy is thermalized and inflates an overpressured cocoon;
the expansion of the cocoon compresses the surrounding medium and, as
long as the energy transferred to the ambient by the expansion work 
is higher than the background energy (even after the jet has ceased its activity), the compression 
of the medium will be mediated by an expanding shock (bow shock) (see Eq. (\ref{eq:mach}))
where the energy transferred to the medium can be dissipated. 
 
A high fraction of the injected energy can be deposited into the ambient medium 
by the expansion work ($\sim 60\%$ at the end of the active phase but even higher
afterward, see Fig. \ref{fig:energyplot}) while only a small fraction is retained in the cocoon 
and in the entrained ambient material ($<10\%$). 
As it has been shown e.g. by Ruszkowski et al. (2003), an expanding
bubble in pressure equilibrium with the surroundings can transfer to
the medium around $40 \%$ of the injected energy but, since mostly in the
active phase of injection the expanding cocoons are strongly overpressured this fraction 
can be much higher in our simulations. 

Once the jet has ceased its activity it quickly loses its collimation
and disappears while the high entropy cocoon formed by the thermalized material
of the jets buoyantly rises into the atmosphere entraining the ambient medium.
As proposed in other works the energy of the rising plumes (or bubbles) associated with the relics of the 
radio lobes can be dissipated thanks to the expansion work done as they rise in the decreasing pressure ambient 
(Roychowdhury et al. 2004), to the mixing between high and low entropy material (e.g. Dalla Vecchia et al. 2004) 
or to the viscous dissipation of sound waves generated by the rising lobes (Ruszkowski et al. 2004) and generally
is high enough to balance cooling losses and limit cooling flows in the cluster cores. 
But since we have shown that the energy retained in the rising plumes is negligible compared to the energy dissipated
at the shock front, in our discussion we neglected the effects due to the entrainment between the high
entropy material of the plumes and the ambient medium. 

The energy dissipated at the shock front has been evaluated looking at the entropy profiles of the shocked ambient 
medium: we found that after the passage of the bow shock the entropy profile as a function of Lagrangian mass 
coordinates tends to reach a steady condition and only small fluctuations due to the action of the rising lobes
are observed, thus confirming that in our simulations the expanding bow shock 
is the main heating mechanism. As it was shown in Fig. \ref{fig:entropyplot} the entropy of the ICM can be 
substantially raised as long as the energy associated with the expanding shock is greater than the energy of the
unperturbed medium contained inside the shock radius: as can be seen in Fig. \ref{fig:entropyplot} (upper solid
curves) and Fig. \ref{fig:entrt} 
only the higher power jets ($5\times10^{46}$ erg s$^{-1}$) can raise the entropy of the central core of  
clusters up to values $>$ 125 $(h/0.7)^{-1/3}$ keV cm$^2$ which represents the ``entropy floor '' determined by 
Lloyd-Davies et al. (2000). Moreover, the data obtained with the higher power jets are in good agreement with 
the average core entropies determined by Ponman et al. (2003) only for the small groups ($T_{\rm ew} <$ 1 keV).

A huge amount of energy is required in order to obtain the observed core entropy with a single energetic event.
Using the same isochoric approximation adopted to calculate the ``equivalent heating'', we estimate that
average energies around 5.3 keV, 4.6 keV and 3 keV per particle for the $T_{\rm ew}=0.5$ keV, 
$T_{\rm ew}=1$ keV and $T_{\rm ew}=2$ cluster respectively are required to increase the average core entropy of
our baseline model 
up to values $>$ 125 $(h/0.7)^{-1/3}$ keV cm$^2$, with a single energy injection at $z=0$. 
In Fig. \ref{fig:tdsplot} we plotted the equivalent amount of energy per particle required to increase the entropy 
of the initial atmosphere up to the entropy determined by the jet injection while   
in Table \ref{table:tdiss} we show for every simulation the average energy per particle dissipated by the jets inside the 
whole computational domain (first value) and inside the cluster core (second value) at the end of our simulations: 
these values are calculated averaging the values of the curves plotted in Fig. \ref{fig:tdsplot} over the
mass contained in the whole computational domain and in the cluster core respectively.
We can see that the energy dissipated by the $5\times10^{46}$ erg s$^{-1}$ jets in the three cluster cores is high enough 
to raise the core entropy up to the entropy floor but nevertheless the lower power cases can give at least in the cluster 
cores the heating energy around 1 keV per particle required in cosmological simulations to satisfy observational 
constraints. 
\begin{table}
\caption{Average energy per particle dissipated by jets of different power $L_{\rm j}$ in the three clusters
characterized by $T_{\rm ew}$=0.5, 1, 2 kev. Reported are the average energy per particle dissipated inside the entire
domain (first value) and inside the cluster core $R < r_{\rm s}$ (second value) at the end of the simulations.}
\begin{center}
\begin{tabular}{cccc}  \hline \hline
$L_{\rm j}$ (erg s$^{-1}$)  & $T_{\rm ew}$=0.5 keV & $T_{\rm ew}$=1 keV & $T_{\rm ew}$=2 keV \\

\hline

$10^{46}$ &  1 - 3.2  keV & 0.5 - 2 keV & 0.2 - 1 keV\\
\hline

$5\times10^{46}$ & 4.1 - 9.7 keV & 2.1 - 5.6 keV & 0.9 - 3.1 keV \\
\hline

\end{tabular}
\end{center}
\label{table:tdiss}
\end{table}
The low power cases can efficiently dissipate their energy inside the
cluster core since at least at the 
beginning of the evolution their cocoons are overpressured with
respect to the ambient medium, thus driving
strong shocks through it. On a larger scale these shocks loose their
strength, and they thus resemble 
observations of sound waves and weak shocks propagating in the ICM
(Fabian et al. 2003, Forman et al. 2003). 

The duration of the overpressured phase during which the jets can
efficiently dissipate their energy is
mainly determined by the ratio between the injected energy and the energy of the unperturbed medium
(see Table \ref{table:cases}) and so it depends
strongly on the characteristics of the medium through which the jets
propagate. As shown by Reynolds et al. (2002) 
even if a jet can efficiently shock the core of a rather shallow entropy profile 
(they assumed an isothermal $\beta$-model with $\beta=0.5$) the lower
entropy material outside the shocked region
will flow inward taking the place of the strongly shocked medium thus
minimizing the effect of the energy dissipation.
On the other hand the steeper entropy profiles that we assumed as
initial condition as representative of 
clusters whose ICM is only heated by gravitational processes are
affected more significantly by the interaction with a supersonic jet. 

The fraction of the injected energy that is dissipated inside the
cluster is plotted in Fig. \ref{fig:tdsevol} for every
simulated case as a function of time. In every case, a high fraction
(up to $75\%$) can be irreversibly dissipated but a few
differences can be noticed: for instance, a rather small dependency of
this efficiency on the ratio $E_{\rm j}/E_{\rm th}$ between the total
injected energy and the thermal energy of the cluster (with $\approx
10\%$ differences). Cocoons
characterized by a higher ratio can remain in an overpressured phase   
for a longer time driving shocks across a higher fraction of the ICM
and dissipating more efficiently their energy.
As noticed before, also less powerful jets can dissipate their energy
onto the ICM, at least at the beginning of their evolution. Therefore, 
many less powerful jets should be able to heat effectively the ICM as
much as one single powerful event.

Taking into account the simulated case characterized by $E_{\rm
j}/E_{\rm th}$ = 10.9 ($L_{\rm j} = 5\times10^{46}$ 
erg s$^{-1}$, $T_{\rm ew} = 0.5$ keV) which is one of the cases that
better satisfies the observational constraints (see Figs.
\ref{fig:lumt} and \ref{fig:entrt}), it is evident that at the $T_{\rm
ew} = 0.5$ keV scale an energy around 8.7 times greater than the
initial thermal energy of the unperturbed cluster (assuming an
efficiency around $80\%$) must be dissipated in order 
to heat the ICM up to the observed values. 
This estimate is consistent with the results obtained by Roychowdhury
et al. (2004) even if they modeled the AGN 
activity in a different way (they considered the work done by a steady
flux of buoyant bubbles): their linear relation 
between the cluster mass and the energy that must be dissipated in the
ICM in order to match the entropy excess at
$0.1R_{200}$ and $R_{500}$ observed by Ponman et al. (2003) is
consistent with our results on the overall energy budget 
at the $T_{\rm ew} = 0.5$ keV scale.
A somewhat different estimate was given by Cavaliere, Lapi \& Menci
(2002): to match the observations of the $L-T$
relation for a $T_{\rm} = 0.5$ keV cluster they require an energy
which is around 3 times the energy of the unperturbed 
medium. They modeled the AGN activity with a self-similar mild
blastwave propagating through the ICM pushed by a piston
which injects energy in the ICM with low power but for long dynamical
timescales: such a blastwave does not efficiently
heat the surrounding medium but tends to push the ICM out of the
virial radius thus decreasing the mass content of the 
cluster and then its X-ray luminosity. 
In our simulations, clusters are heated more efficiently, and the
impact of this on the $L-T$ relation can be evaluated.
Heated clusters tend to expand and cool down adiabatically: when they
recover the gravitational equilibrium 
they have a less dense core and a temperature only slightly higher
than the initial one thus decreasing their
X-ray luminosity. As it is shown in Fig. \ref{fig:lumt} the cases that
are heated up to the observed values 
(Fig. \ref{fig:entrt}) can expand to a new equilibrium to give X-ray
luminosities which are consistent with the observed $L-T$ relation.

From the above discussion, our conclusion is that the mechanism
discussed here can represent one viable way to induce a
non--gravitational heating of the ICM. An extra heating of $\approx
1$ keV per particle, is known to be sufficient for bringing the 
theoretical $L_{\rm X}-T$ relation in agreement with observations 
(see e.g. Borgani et al. 2002). 
We have shown how such a heating level can be provided by AGN jets 
having kinetic power in the range $10^{46}-5 \times 10^{46}$ erg s$^{-1}$. 

The situation of one single, powerful jet which heats the ICM to the
observed level can be quite rare in reality. More probably, several,
less powerful events contribute to the heating. We showed that
the contribution of such events, while smaller, have a non
negligible effect. Other non gravitational mechanism, like the 
``effervescent heating'' or other interactions between AGN-originated
bubbles or plumes of gas and the ICM, could also be simultaneously in
action, and we regard them as alternative, and not exclusive, respect
to the one described here.

Another important detail is that the calculation reported here are
based on the ICM physical characteristics {\it at a redshift $z=0$},
while most probably a jet heating would occur at higher redshifts,
in the environment given by proto--cluster structures.
One could speculate that, depending on the actual number and power of
heating events which happened in these proto--clusters, the final
excess heating can significantly vary among different clusters, thus
explaining not only the break in the self--similarity, but also the
observed spread in the X--ray properties at a given temperature. For
instance, from Fig. \ref{fig:lumt} one can see that, at a temperature
$T \approx 0.7-0.9$ keV, the observed luminosity can vary by more than an
order of magnitude. If the extra--heating of the clusters atmospheres
is due to the AGN activity, and if such activity is somehow linked
with the proto--cluster ``cosmological'' environment, the spread could
be reminiscent of the formation history of the cluster.

To put on firmer ground such speculations, we plan to repeat the
current analysis using several atmospheres of proto--cluster structures taken
from N-Body/SPH simulations at high redshifts. This will allow us to
quantitatively asses the contribution of AGN jets heating on the gas
which will form the clusters atmospheres at redshift $z=0$, and
possibly to follow the heated gas during the cluster formation
process, again with the use of N-Body/SPH simulations.
This phase is preliminary to the attempt to give a parametrization of
the effect of AGN activity on the ICM, or at least of the contribution
due to the jets heating, which could be directly used in
self--consistent cosmological simulations.

\begin{acknowledgements}
The authors acknowledge the italian MIUR for financial support, grant
No. 2001-028773. The numerical calculations have been performed at CINECA
in Bologna, Italy, thanks to the support of INAF.
We acknowledge L. Tornatore and S. Borgani for useful discussions and for
providing the N-Body simulations used in this paper.
\end{acknowledgements}

\end{document}